\documentclass[preprint]{aastex}
\usepackage{longtable}
\usepackage{multicol, nicefrac, enumitem, multirow, units, sidecap, xspace}
\usepackage{fixltx2e}
\usepackage{graphicx}
\usepackage{verbatim}
\usepackage{capt-of}
\usepackage{float}
\usepackage{grffile}
\usepackage{color}

\makeatletter
\def\LT@makecaption#1#2#3{%
  \LT@mcol\LT@cols c{\hbox to\z@{\hss\parbox[t]\LTcapwidth{%
    \sbox\@tempboxa{#1{#2 }#3}%
    \ifdim\wd\@tempboxa>\hsize
      #1{#2 }#3%
    \else
      \hbox to\hsize{\hfil\box\@tempboxa\hfil}%
    \fi
    \endgraf\vskip\baselineskip}%
  \hss}}}
\makeatother

\slugcomment{To Appear in the Astrophysical Journal}
\shorttitle{Hierarchical Formation in Action}
\shortauthors{Zucker et al.}
\newcommand{\spit}{{\em Spitzer}\xspace}
\newcommand{\wise}{WISE\xspace}

\begin{document}
\title{Hierarchical Formation in Action: Characterizing Accelerated Galaxy Evolution in Compact Groups Using Whole-Sky WISE Data\\}

\author{Catherine Zucker}
\affil{Department of Astronomy, University of Virginia,
    Charlottesville, VA, 22904}
\affil{Harvard-Smithsonian Center for Astrophysics,
    Cambridge, MA, 02138}
\author{Lisa May Walker}
\affil{Department of Astronomy, University of Arizona,
    Tucson, AZ, 85721}
\author{Kelsey Johnson}
\affil{Department of Astronomy, University of Virginia,
    Charlottesville, VA, 22904}
\author{Sarah Gallagher}
\affil{Department of Physics and Astronomy, University of Western Ontario, London, Ontario N6A 3K7 Canada}
\affil{Centre for Planetary and Space Exploration, University of Western Ontario, London, Ontario N6A 3K7 Canada}
\author{Katherine Alatalo}
\affil{Hubble Fellow, Carnegie Observatories, Pasadena, CA, 91101}
\affil{Infrared Processing and Analysis Center, California Institute of Technology, Pasadena, California, 91125}
\author{Panayiotis Tzanavaris}
\affil{Laboratory for X-Ray Astrophysics, NASA Goddard Spaceflight Center, Mail Code 662, Greenbelt, MD, 20771, USA}
\affil{CRESST, University of Maryland Baltimore County, 1000 Hilltop Circle, Baltimore MD 21250 USA}
\affil{Department of Physics and Astronomy, The Johns Hopkins University, Baltimore, MD 21218, USA}

\email{catherine.zucker@cfa.harvard.edu}

\begin{abstract}
Compact groups provide an environment to study the growth of galaxies amid multiple prolonged interactions. With their dense galaxy concentrations and relatively low velocity dispersions, compact groups mimic the conditions of hierarchical galaxy assembly. Compact group galaxies are known to show a bimodality in \textit{Spitzer} IRAC infrared colorspace: galaxies are preferentially either quiescent with low specific star formation rates, or are prolifically forming stars---galaxies with moderate levels of specific star formation are rare. Previous \textit{Spitzer} IRAC studies identifying this ``canyon" have been limited by small number statistics. We utilize whole-sky WISE data to study 163 compact groups, thereby tripling our previous sample and including more galaxies with intermediate mid-IR colors indicative of moderate specific star formation rates (SSFRs). We define a distinct WISE mid-IR color-space ($\log[{\frac{\rm f_{12}}{\rm f_{4.6}}}]$ vs. $\log[{\frac{\rm f_{22}}{\rm f_{3.4}}}]$) that we use to identify canyon galaxies from the larger sample. We confirm that compact group galaxies show a bimodal distribution in the mid-infrared and identify 37 canyon galaxies with reliable photometry and intermediate mid-IR colors. Morphologically, we find that the canyon harbors a large population of both Sa-Sbc and E/S0 type galaxies, and that they fall on the optical red sequence rather than the green valley. Finally, we provide a catalog of WISE photometry for 567 of 652 galaxies selected from the sample of 163 compact groups. \end{abstract}

\section{Introduction}
With their high number densities, low velocity dispersions, and frequent reservoirs of intragroup diffuse gas \citep{hickson92}, compact groups are dense associations of three or more bright galaxies that routinely undergo prolonged gravitational interactions \citep{hickson92}. Due to these properties, they bear a strong resemblance to conditions in the earlier universe, making them an ideal laboratory for studying physical processes such as the formation of central supermassive black holes, galaxy clusters, super-star clusters, and dwarf galaxies experiencing multiple interactions \citep{konstantopoulos13, tzanavaris14}. Compact groups also play a critical role in understanding galaxy evolution, as their frequent gravitational interactions affect the gas processing and specific star formation rates (SSFR; star formation rate per stellar mass) of member galaxies. They also tend to be more difficult to characterize than either galaxy clusters or merging pairs: unlike virialized galaxy clusters with thousands of members, compact groups may be more sensitive to the initial configuration of the handful group members, and, unlike galaxy pair mergers, initial conditions in compact groups are more difficult to constrain with models due to a large number of free parameters. 

In order to assess the impact of the compact group environment on star formation and galaxy evolution, \citet{johnson07} studied the mid-infrared colors of these galaxies and discovered an underdensity between compact group galaxies that are actively star-forming and those that are relatively quiescent. The underdensity in colorspace suggests a rapid transformation of galaxy properties, in the sense that this dearth in \textit{Spitzer} IRAC colorspace ($\log[{\frac{\rm f_{8.0}}{\rm f_{4.5}}}]$ vs. $\log[{\frac{\rm f_{5.8}}{\rm f_{3.6}}}]$) is a transition region between spirals and bulge-dominated galaxies \citep{walker10}. This transition region, or IRAC ``canyon", is occupied by relatively few galaxies with moderate SSFRs and not seen in comparison samples of field galaxies, interacting pairs, or the center of the Coma Cluster \citep{walker10,walker12}. The Coma infall region is the only comparison sample that shows a similar distribution to compact group galaxies in IRAC colorspace \citep{walker12}. This is consistent with the idea that the infall region best mimics the properties of the compact group environment, as they both have high galaxy densities and reservoirs of cold gas. Finally, in addition to showing a bimodality in mid-IR colorspace, \citet{tzanavaris10} also find a significant gap between low ($\le 4.3 \times 10^{-12}\; \rm{yr}^{-1}$) and high ($\ge 1.6 \times 10^{-11}\; \rm{yr}^{-1}$) SSFR systems in the compact group environment. The recent work of Lenkic et al. (2016, MNRAS, Submitted) further supports this SSFR bimodality. The lack of compact group galaxies with moderate levels of specific star formation, in contrast to other environments, implies that {\it compact group galaxies experience accelerated evolution}. Star formation enhancement due to gravitational interactions (including mergers) of galaxies in this dense environment likely also plays a role in this observed bimodality.

To further investigate the nature of compact group galaxies, \citet{cluver13} used the {\em Spitzer} Infrared Spectrograph (IRS) to study the PAH and $\rm {H_2}$ lines in the systems.  The first compact group studied in detail in this way was Stephan's Quintet \citep{appleton06,cluver10}, where enhanced warm H$_2$ emission was seen in quantities that could not be explained by photoionization from star-forming regions alone. This emission was associated with the large bow-shock seen between the intruder (NGC\,7319) and the interacting pair (NGC\,7318a \& 7318b).  \citet{cluver13} followed up this effort, investigating the mid-IR spectral properties of 74 galaxies in 23 Hickson Compact Groups, including 17 galaxies within the original IR gap of \citet{johnson07}.  These results show that the galaxies that are in the IR gap are the most likely to show enhanced warm H$_2$ from \textit{Spitzer}, and thus are considered galaxies most likely to have shock-enhanced interstellar media, hinting that shocks in these systems might be a significant contributor to canyon galaxies' rapid evolution.

All previous IR-based studies --- using data from the \textit{Spitzer Space Telescope} -- are limited by their small sample sizes. \citet{johnson07} studied only 12 compact groups (45 galaxies), while \citet{walker12} only studied 49 compact groups (174 galaxies). The largest sample \citep{walker12} yielded only five galaxies located in the mid-infrared canyon region, a minuscule number when one seeks to understand the properties of these unique galaxies and their evolution. The small sample sizes of previous studies are due to both limited sky coverage and the narrow imaging bands of \spit, which forces a redshift cut as the polycyclic aromatic hydrocarbon (PAH) features shift out of the $8 \micron$ band at  $z=0.035$. In contrast, the Wide-field Infrared Survey Explorer (\wise) telescope has whole-sky coverage and bands wide enough that the PAH features do not shift out of the $12 \micron$ band until $z=0.52$.  Using \wise data, we perform analyses similar to those of \citet{johnson07} and \citet{walker12}. We expand the sample selection to a total of 163 compact groups (652 member galaxies), thereby drawing from all 49 groups from \citet{walker12}, plus an additional 114 groups, increasing the number of compact groups with mid-IR analyses by a factor of three. In doing so, we are able to identify additional galaxies in the mid-infrared canyon region, producing a sizable sample of moderate SSFR galaxies that will enable a more comprehensive analysis of the processes that influence galaxy evolution in this environment.

\section{Data}
\subsection{Sample}

For this study, we draw our sample from groups in the Hickson Compact Group catalog \citep[HCG,][]{hickson82} and the Redshift Survey Compact Group catalog \citep[RSCG,][]{barton96}. Since our sample is culled from two different catalogs, it is necessary to consider their differing selection criteria. The original 100 Hickson Compact Groups were identified through a systematic visual search of the Palomar Observatory Sky Survey prints. In the original sample, each HCG had to contain at least four galaxies (within three magnitudes of the brightest galaxy) and had to satisfy isolation and compactness criteria---requirements that excluded the addition of galaxy clusters in the catalog.  Follow-up radial velocity data indicated that several groups had only three members within 1000 km s$^{-1}$ of the median group velocity, but these groups were not rejected \citep{hickson92}. Seven other groups (HCGs 9, 11, 18, 36, 41, 77, 78) were shown to have less than three accordant members, and these groups are excluded from our analysis \citep{hickson92}. Following the convention from \citet{hickson92}, we also exclude discordant galaxies that by projection appear to be members of accordant groups. 

In contrast to the HCGs, the 89 RSCGs were chosen from a complete, magnitude-limited redshift survey, using a friends-of-friends algorithm, with parameters chosen to identify compact groups similar to the HCGs \citep{barton96}. ``Neighbors'' were identified based on projected separation ($\Delta d < 50 $ kpc) and median radial velocity difference ($\Delta v <1000$ km s$^{-1}$). The radial velocity criterion is concordant with Hickson's procedure of rejecting group members with a median line-of sight velocity difference $>1000 $ km s$^{-1}$ \citep{hickson92}. Unlike Hickson, \citet{barton96} did not take galaxy magnitudes into account, nor did they implement Hickson's isolation criterion. As a result, some RSCGs are embedded in clusters, with the most extreme examples being RSCGs 67 and 68 (embedded in the Coma Cluster), RSCG21 (embedded in the Perseus Cluster), and RSCG11 (embedded in the core of Abell cluster 194). It is important to recognize that the cluster environment acts as a confounding variable in our analysis of the compact group environment, so we have excluded the most extreme examples---RSCGs 11, 21, 67, and 68---from our analysis. Several other RSCGs are located on the edge of clusters, including RSCG12 on the outskirts of the Zwicky cluster \citep{barton98}, or RSCG65 on the outskirts of the Virgo cluster \citep{barton96}. As the cluster outskirt environment is expected to be less of an influence than the cluster core environment (being the only other comparison sample from \citet{walker10} to exhibit a gap), we include these groups in our analysis. Finally, though the HCG and RSCG catalogs had different selection criteria, there are fifteen RSCGs that are either HCGs or part of HCGs. In these instances, we only use the HCG group to avoid double-counting. Excluding all discordant HCGs (7 groups) and RSCGs embedded near the core of clusters (4 groups), our final tally of distinct compact groups is 163 (93 HCGs and 70 RSCGs), totaling 652 constituent galaxies. 

\subsection{WISE Photometry}
The original ALLWISE atlas images are intentionally blurred, with their PSF profiles having FWHM values which are $\sqrt2$ times larger than the single exposure values. As such, we utilize new ALLWISE coadds\footnote{http://unwise.me/} from \citet{lang14}, which preserve the native resolution of the raw frames ($\rm \approx 6.1\arcsec, 6.4\arcsec, 6.5\arcsec, \; and \; 12.0\arcsec$ for bands W1, W2, W3, and W4). We convert to Janskys using $f_0\times10^{-0.4 \rm MAGZP}$, taken from Section 4.3 in the explanatory supplement for the ALLWISE image atlas. We adopt the same zero magnitude flux densities ($f_0=$306.682, 170.663, 29.0448, and 8.2839 for W1, W2, W3, and W4) implemented in the conversion of ALLWISE source catalog raw fluxes to magnitudes, which assumes a $f_\nu^{-2}$ power law spectra. The zero-point value MAGZP equals 22.5 for all four bands, as \citet{lang14} scale the reprocessed ALLWISE atlas images to this zero point. To ensure a common pixel scale and resolution, we convolve the W1 ($3.4\micron$), W2 ($4.6\micron$), and W3 ($12\micron$) images to the W3 PSF, using the WISE ``frame" kernels from \citet{aniano11}. We avoid convolving to the W4 ($22\micron$) PSF, as this band has low sensitivity and will only add noise for a majority of the galaxies. Instead, we derive an independent aperture for galaxies with high signal-to-noise in W4, and pull the rest of the photometry from the ALLWISE source catalog for unresolved or low signal-to-noise systems.

We perform photometry on the convolved W1, W2, and W3 images and select W4 images using SURPHOT, a multi-wavelength photometry code which measures flux densities within a single aperture across multiple wavebands \citep{reines08}. We use SURPHOT to detect a contour level (1-3$\sigma$ above the background) in a specified region of a reference image---an averaged, weighted ($\lambda ^{-1}$) image of the W1,W2, and W3 bands (hereafter referred to as W123). As the W123 reference image is weighted by $\lambda ^{-1}$, W1 drives the aperture, followed by W2 and W3. Once an aperture is determined using the reference image, SURPHOT assigns this reference aperture identically to the first three WISE bands. Three sets of background annuli are also applied by expanding the aperture in W123 by two specified factors ($2.0-2.5\times, 2.0-3.0\times, 2.5-3.0\times$), which are then applied to the W1, W2, and W3 bands. For each annulus ($2.0-2.5\times, 2.0-3.0\times, 2.5-3.0\times$), the background flux is calculated using both the mode and the resistant mean, which reduces the effect of neighboring galaxies contaminating the background annulus. These local background fluxes are then subtracted from the source flux. Uncertainties are calculated for all SURPHOT fluxes by determining the standard deviation in the set of fluxes due to differences in the background. As the background flux is calculated in six ways (mode and the resistant mean for the three types of background annuli), we determine the standard deviation in the set of six background-subtracted source fluxes.

Thus, we divide the photometry into three groups, based on their resolution and signal-to-noise ratio in W1 through W4:
\begin{enumerate}

\item{For galaxies resolved in W1, W2, W3, and W4, we perform photometry on W1, W2, and W3 with SURPHOT, applying a matched aperture derived from W123 to each of the first three convolved images. We then use SURPHOT to define an independent aperture using only the W4 image. To determine whether a source is resolved in W4, we use the reduced $\chi^2$ of the W4 profile-fit photometry measurement (``w4rchi2") from the ALLWISE source catalog. This parameter indicates the goodness of fit between the source and the PSF, with resolved and extended galaxies exhibiting larger $\chi^2$. We flag any galaxy with $\rm w4rchi2>2$ as a resolved and extended source. Of the 652 galaxies in our sample, 59 galaxies are individually resolved in the W123 reference image and also have $\rm w4rchi2\ge2$, so SURPHOT is used to calculate fluxes in all four bands.}

\item{For galaxies resolved in W1, W2, and W3, but with a $\rm w4rchi2<2$, we perform photometry on W1, W2, and W3 using the same methodology as above and pull the W4 fluxes (w4mpro) from the ALLWISE source catalog.}

\item{For galaxies that a) have too low signal-to-noise for us to apply a $1\sigma$ contour to them and b) cannot be individually resolved with a $3\sigma$ contour due to small projected separation, we pull photometry for W1, W2, W3, and W4 (w1mpro,w2mpro,w3mpro,w4mpro) from the ALLWISE source catalog} 
\end{enumerate}

Star subtraction is done on all SURPHOT galaxies where a confirmed star is visible within the galaxy aperture (15 galaxies). We use the stellar-dominated band with the larger PSF (W2) to define a circular aperture around the star, and this is then subtracted from the W1 through W3 fluxes. No star-subtracted galaxies are resolved in W4, so the W4 fluxes for these 15 galaxies are pulled from the ALLWISE catalog.

As the WISE absolute photometric calibration is based on profile-fitting of point sources, we also apply a flux correction to account for the extended nature of our galaxies. \citet{jarrett13} gives the corrections in magnitudes: $+0.034$, $+0.041$, $-0.030$, $+0.029$, for the $3.4\micron,4.6\micron,12 \micron, 22\micron$ bands, respectively. We convert from a magnitude correction to a flux correction using the relationship $\rm m_{apcor}=-2.5\log_{10}f_{apcor}$, taken from section 2.3 of the WISE preliminary explanatory supplement.

Though most groups are local, there are several with redshifts exceeding 0.1. For the sake of consistency, we apply a k-correction to all galaxies using the low resolution spectral templates and FORTRAN fitting code from \citet{assef10}. \citet{assef10} determine low resolution empirical SED templates from the near-ultraviolet (0.03 $\micron$) to the mid-infrared (30 $\micron$) that accurately reproduce galaxy properties. Their methodology is based upon the assumption that every galaxy SED is some linear combination of old stellar populations, continuously star-forming populations, starbursting populations, and post-starbursting populations. We apply the \citet{assef10} fitting routine to our WISE W1 through W4 fluxes, using the spectral response curves from \citet{wright10}. We note that for a small subset of the sample ($\approx15 \%$ of the full sample) the fitting routine returns a high reduced $\chi_\nu^2$ value ($\chi_\nu^2 >150$); this occurs preferentially for galaxies with high signal-to-noise in W4, to which we independently fit an aperture using SURPHOT. None of the galaxies with poor $\chi_\nu^2$ values are in high redshift groups, and we expect this effect to be negligible.  A majority of the galaxies in our full sample yield a $\chi_\nu^2 \lesssim 1 $.

Finally, 77 of the 652 galaxies are either not identified in the ALLWISE source catalog (to within $5\arcsec$) or have a negative flux measurement provided by the catalog, so these galaxies are excluded from the sample. We also impose a signal-to-noise cutoff of 2 in all four bands, as the WISE profile-fit magnitude is substituted with a $2\sigma$ brightness upper limit when the source fails to meet this signal-to-noise cutoff. Providing that the galaxy has an upper limit in only a single band, we include galaxies with $0 < S/N <2$, but mark them as such in all color-color plots. Our final sample of galaxies, including those with upper limits in one band, is 567 galaxies (428 with reliable photometry and 139 with upper limits).

\subsection {SDSS Photometry}
In \S3.5, we plot the distribution of a subsample of our galaxies in SDSS colorspace, and we derive our $u,g,r,i$ photometry for this analysis from the DR12 SDSS photometric catalog.\footnote{http://skyserver.sdss.org/dr12/en/tools/crossid/crossid.aspx} We use model magnitudes (``modelMag"), which implement the better of two fits (de Vaucouleurs and exponential model) in the \textit{r}-band to derive a matched aperture and determine the flux through all bands; modelMags are recommended for calculating the colors of extended objects.\footnote{https://www.sdss3.org/dr10/algorithms/magnitudes.php\#mag\_model}.  We apply standard Galactic extinction corrections in magnitudes following \citet{schlegel98}. While we use the \citet{assef10} code to correct the mid-infrared fluxes,  we k-correct our SDSS magnitudes using the IDL routine calc\_kcor.pro from \citet{chilingarian11}.

\section{Results and Discussion}
\subsection{Comparison of IRAC and WISE Colorspaces}
Since the available WISE bands [$3.4\micron$ (W1), $4.6\micron$ (W2), $12\micron$ (W3), $22\micron$ (W4)] are different from the Spitzer IRAC bands [$3.6\micron$ (IRAC1), $4.5\micron$ (IRAC2), $5.8\micron$ (IRAC3), $8.0\micron$ (IRAC4)], we can not directly map \textit{Spitzer} IRAC colors to WISE colors. In creating a distinct region of WISE color-color space, we need to be cognizant of the similarities and differences between the two instruments. As shown in \citet{jarrett11}, the passbands for W1 and W2 closely correspond to the passbands for IRAC1 and IRAC2, though W1 tends to be slightly bluer with respect to IRAC1, while W2 is slightly redder with respect to IRAC2 \citep[see Figure 1 from][]{jarrett11}, though this should have minimal impact on our results. Moreover, W3 has almost no overlap with IRAC4, and the W3 band should be more sensitive to $11.3\micron$ PAH emission and $10\micron$ silicate absorption in nearby star-forming galaxies, though it will be less sensitive to the stronger PAH features at $6.2\micron$ and $7.7\micron$ \citep{jarrett11}. Thus, a) both W1 and W2 and IRAC1 and IRAC2 are dominated by stellar photospheric emission while b) W3 will be more sensitive than IRAC4 to $11.3\micron$ PAH emission in nearby galaxies and c) W4 has no IRAC comparison band (it is closest to the $24\micron$ band on MIPS), and will have large contributions from relatively warm thermal dust emission, if present. 

\begin{figure}[H]
\centering
\includegraphics[width=.45\textwidth]{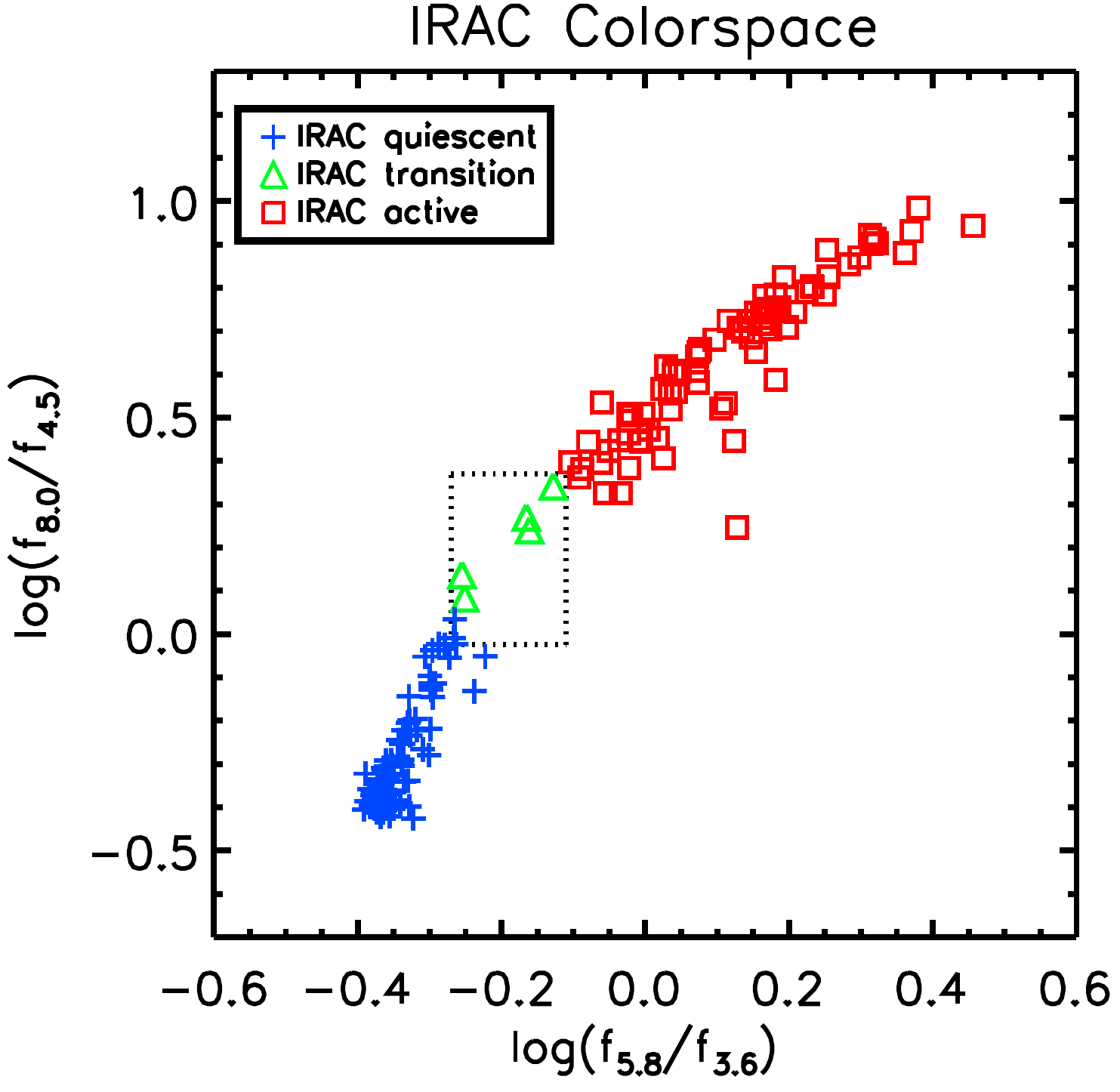}
\includegraphics[width=.44\textwidth]{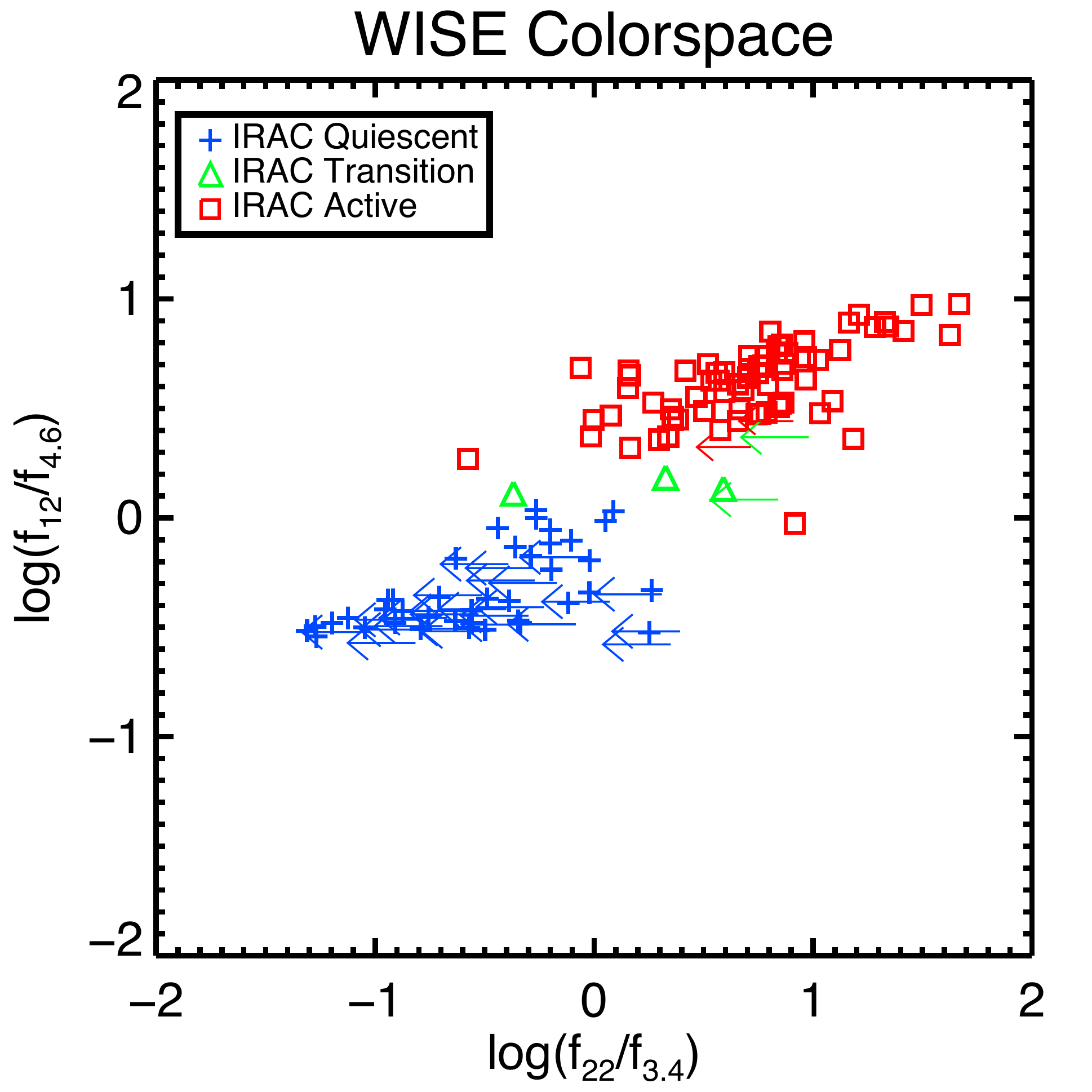}
\caption{\textit{Left:} IRAC colorspace distribution of the sample of 174 galaxies from \citet{walker12}. \textit{Right:} WISE colorspace distribution of all the galaxies common to both the \citet{walker12} sample and our full sample (141 galaxies). The galaxies in WISE colorspace are color-coded by their location in IRAC colorspace.}
\label{fig:colorspace_comp}
\end{figure}

Since different mid-infrared bands are probes of different emission mechanisms, we can use different color combinations to separate galaxies that are actively star forming from those that are relatively quiescent. Using IRAC colors, \citet{walker12} are able to separate galaxies dominated by stellar light from galaxies dominated by PAH and thermal dust emission, by plotting $\log[{\frac{\rm f_{8.0}}{\rm f_{4.5}}}]$ vs. $\log[{\frac{\rm f_{5.8}}{\rm f_{3.6}}}]$. This is shown in the left panel of Figure \ref{fig:colorspace_comp}. For the 141 galaxies in our full sample that we have in common with the sample of \citet{walker12}, we are able to reproduce this same trend by plotting $\log[{\frac{\rm f_{12}}{\rm f_{4.6}}}]$ vs. $\log[{\frac{\rm f_{22}}{\rm f_{3.4}}}]$, as shown in the right panel of Figure \ref{fig:colorspace_comp}. Unfortunately, the two bounding canyon galaxies from the \citet{walker12} sample (HCG37e \& HCG57h) have a low signal-to-noise in the $22\micron$ band, so we are not able to directly map the IRAC canyon to the WISE canyon. Instead, we use the full sample to statistically define the bimodality in \S3.2. 

Comparing the two distributions, it appears that the spread of galaxies in WISE colorspace shows more scatter than their spread in IRAC colorspace. We attribute this in part to larger photometric errors induced by the introduction of the lower resolution and lower sensitivity WISE $22 \micron$ and $12\micron$ bands and the smaller aperture of the WISE instrument. In addition to larger photometric errors, we hypothesize that the WISE canyon could potentially be less prominent than the \textit{Spitzer} canyon due in part to the inclusion of these higher wavelength bands, which probe a different regime than the $3-8\micron$ IRAC bands. It is generally assumed that $6-9\micron$ features (traced by IRAC4), will be stronger in ionized PAHs (PAH$^+$), while the $11.3\micron$ feature (traced by W3) will be stronger in neutral PAHs (PAH$^{0}$) \citep{galliano08, wu10, sadjadi15}. As such, the UV-excited $7.7\micron$ feature preferentially traces current star formation, while the $11.3\micron$ feature preferentially traces the diffuse ISM, and is often found in early-type galaxies with low star formation rates (see \citet{cluver13} and references therein). Moreover, there is no IRAC comparison band for W4, which traces relatively warm thermal dust emission. 

Along this line of thought, \citet{alatalo14} show an ``elbow" between late and early type galaxies in WISE [W1-W2] vs. [W2-W3] colorspace and hypothesize that this result is due to the fact that galaxies transition in optical colors (quenching star formation) before IR color (shedding their interstellar media, as traced by the thermal dust $22\micron$ emission and 11.3$\mu$m neutral PAH feature), a finding also suggested in \citet{walker13}. If the WISE canyon seen in compact groups is also caused in part by the shedding of the galaxies' ISM, this could explain why the canyon is less prominent in WISE colorspace than in \textit{Spitzer} color space. Since no IRAC bands adequately trace the $22 \micron$ thermal dust emission or the $11.3$ micron neutral PAH emission, ISM levels would not significantly contaminate the distribution of galaxies, potentially producing a cleaner spread in IRAC colorspace.

\subsection{WISE Colorspace Distribution for Full Sample}
\begin{figure}[H]
\centering
\includegraphics[width=\textwidth]{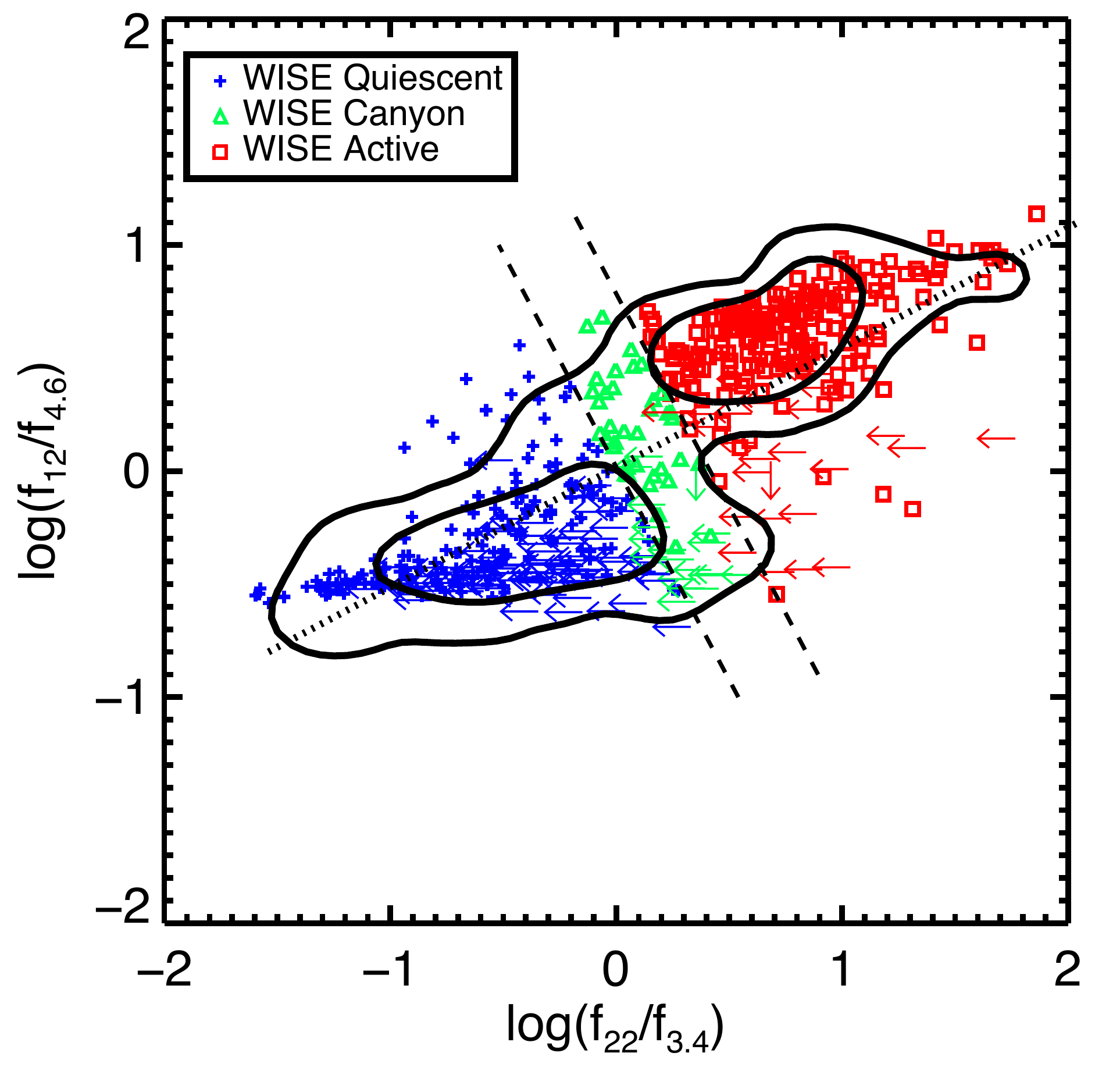}
\caption{WISE colorspace distribution of our full sample. We overlay isodensity contours ($n=5.69, 20.01$ galaxies per bin) on to our full sample, along with a line-of-best fit (dotted line) and the canyon bounds (dashed lines). We color-code galaxies by their location in WISE color-color space. We fit a canyon to the full sample by a) determining the minimum contour level that separates the active and quiescent populations and b) defining these regions as those lying at least $2\sigma$ above this density threshold. We determine a line-of-best fit for the full sample, and then draw lines perpendicular to this fit and intersecting the $2\sigma$ contours at the points where the distance between the two contours is at a minimum.} 
\label{fig:colorspace_fullsamp}
\end{figure}

Having demonstrated the bimodality of the \citet{walker12} \textit{Spitzer} sample in WISE color-color space, we plot the distribution of our final sample in Figure \ref{fig:colorspace_fullsamp}. Though less pronounced than in \citet{walker12} and \citet{johnson07}, there is still a visible underdensity of galaxies apparent in both samples. We note that galaxies with low signal-to-noise in W4 potentially contaminate the WISE canyon at low $\log[{\frac{\rm f_{12}}{\rm f_{4.6}}}]$ color, and that these galaxies would likely shift leftward towards the quiescent population with more reliable photometry. This leftward shift would then further demarcate the canyon region, producing a greater resemblance to the bimodality in \citet{walker12}. We emphasize that W4 only achieves a $5\sigma$ point source sensitivity of $\approx6$ mJy, so this is expected to have a tangible effect on those systems with minimal emission at $22 \micron$.

To define the location of the WISE canyon, we fit isodensity contours to the full sample. We determine the two-dimensional density function of the $\log[{\frac{\rm f_{12}}{\rm f_{4.6}}}]$ and $\log[{\frac{\rm f_{22}}{\rm f_{3.4}}}]$ colors using the IDL routine hist\_2d.pro, adopting a bin size of 0.3 x 0.3. To provide some smoothing, we interpolate the image to a grid $5\times$ smaller by using the IDL routine min\_curve\_surf.pro. Assuming Poisson statistics, we determine the minimum contour level that separates the active and quiescent populations (n=12.85 galaxies per bin). We then define the bulk of the active and quiescent populations as regions lying at least $2\sigma$ above this density threshold. With a Poissonian distribution, $\sigma=\sqrt{\mu}$=3.58 galaxies per bin, so the $2\sigma$ contours above our minimum threshold lie at n=20.01 galaxies per bin. We determine a line-of-best fit for the full sample, and then draw lines perpendicular to this fit and intersecting the $2\sigma$ contours at the points where the distance between the two contours is at a minimum. The region bounded by these two perpendicular lines becomes our WISE canyon. 

In Figure \ref{fig:colorspace_fullsamp} we overlay the contours which lie $\pm2\sigma$ (n=5.69, and n=20.01 galaxies per bin) from the minimum contour level separating our active and quiescent regions (n=12.85 galaxies per bin). We plot the line-of-best fit using the dotted line and the canyon boundaries with the perpendicular dashed lines. Using this canyon definition, we color-code the galaxies as either ``quiescent", ``canyon", or ``active". Of the 567 galaxies in our full sample, 57 (10\%) fall in the WISE canyon compared to only 5 galaxies (3\%) from the \citet{walker12} ``rotated'' sample. \textit{We emphasize that a significant fraction (20 of 57) canyon galaxies have unreliable photometry in W4, and would likely become ``quiescent" with more reliable data, so we focus much of our analysis on the 37 canyon galaxies with high signal-to-noise in all four bands}.  A list of results and properties for all galaxies in our full sample is shown in the Appendix.

\begin{figure}[H]
\centering
\includegraphics[width=.45\textwidth]{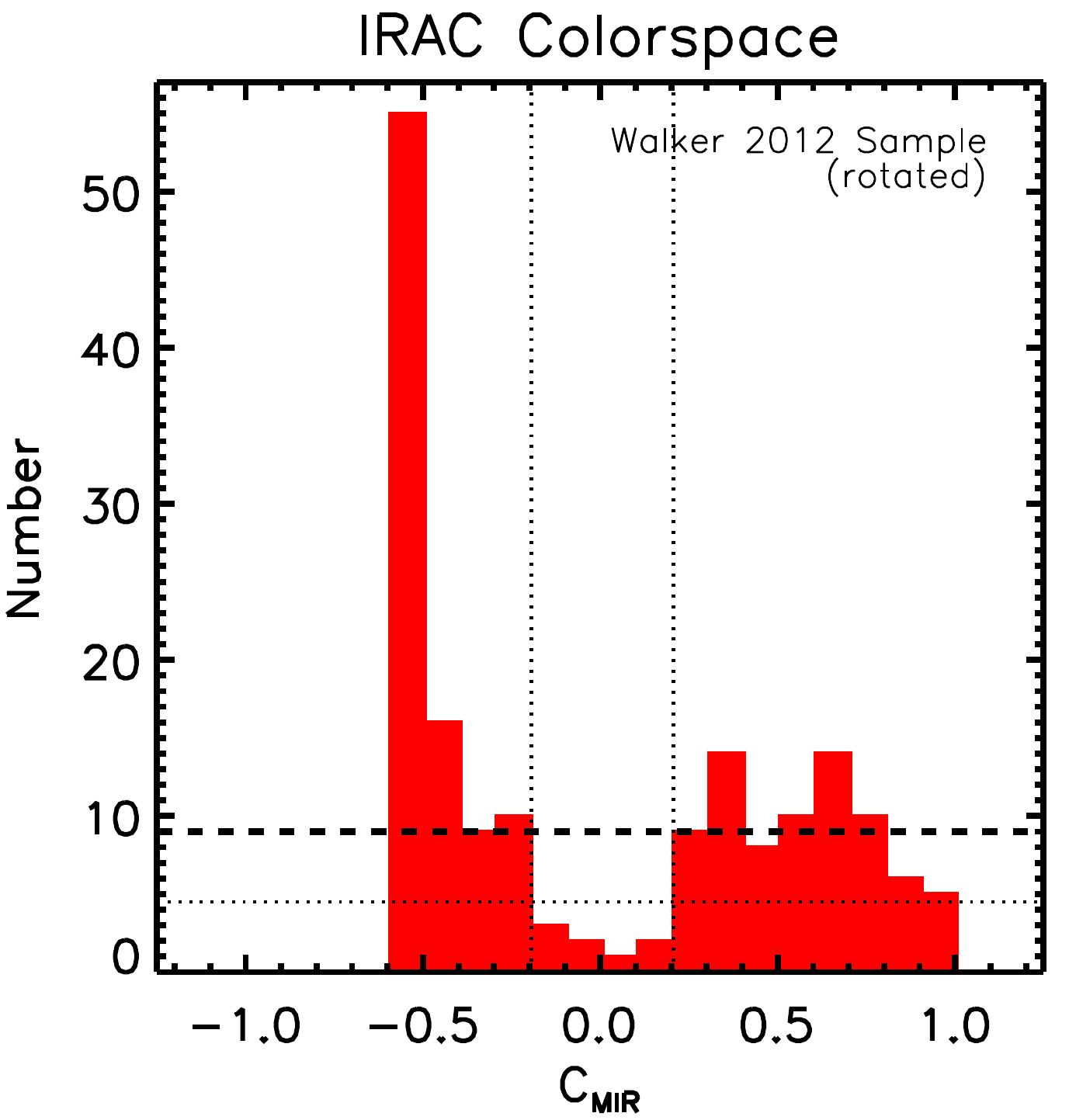}\\
\includegraphics[width=.5\textwidth]{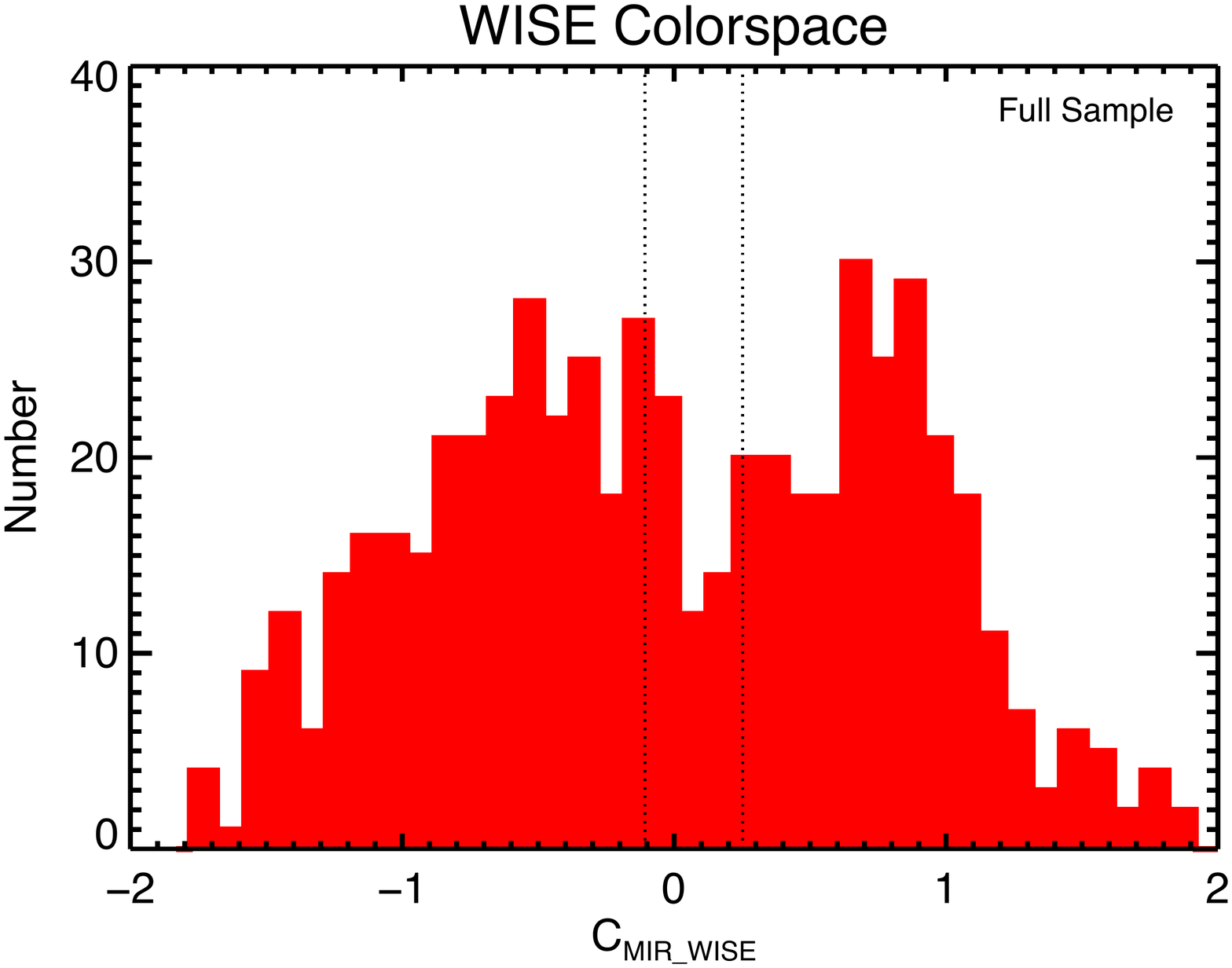}\\
\caption{Rotated colorspace distribution of the \citet{walker12} sample (top), along with our full sample (bottom). In each case, a straight line fit to the colorspace distribution of the corresponding sample became the axis. For the \citet{walker12} IRAC sample, the dashed line indicates the median number per bin, with the horizontal dotted line half that value; the vertical dotted lines indicate the boundaries of the canyon, defined to be where the distribution is less than half the median value. Though we do not use the histogram to define the canyon, we also indicate the region corresponding to the new WISE canyon with vertical dotted lines.}
\label{fig:histograms}
\end{figure}

\citet{walker12} consider the mid-IR colors of their sample in two ways: rotated so that a straight line fit to the colorspace distribution (left panel of Figure \ref{fig:colorspace_comp}) becomes the axis, and unwrapped so that a curve fit to the colorspace distribution becomes the axis. In both cases, the sample is shifted to a mean value of zero. As seen in the top panel of Figure \ref{fig:histograms}, they define the canyon to be where the histogram is less than half its median value; the difference in the canyon bounds produced by the two methods was not statistically significant. For this work, we consider only the rotated distribution, as our WISE colorspace distribution lacks the curvature of the IRAC distribution. Figure \ref{fig:histograms} shows the rotated mid-IR colors for the \citet{walker12} sample (top panel) alongside our full sample (bottom panel); we mark the boundaries of the canyon in both samples with the vertical dotted lines. We see that our WISE canyon, applied in two-dimensions, agrees fairly well with the underdensity apparent in the one-dimensional distribution. This underdensity could potentially become more pronounced with more reliable photometry, as the upper limit canyon galaxies, most of which lie in the $\rm C_{MIR\_WISE}$ bin centered at -0.05, would shift towards more negative $\rm C_{MIR\_WISE}$ values. We also see that, although both IRAC and WISE show a bimodal distribution, our WISE sample is broadened in $\rm C_{MIR\_WISE}$ when compared to IRAC, which makes sense given that WISE probes a much broader range of ISM parameter space. 

\subsection{Star Formation Rate and Stellar Mass}
The WISE bands have proven to be excellent probes of both star formation rate and total stellar mass content, two critical parameters needed to understand the formation and evolution of galaxies. Evolved stellar populations emit the majority of their light at near-infrared wavelengths ($1-5\micron$), so both the (W1) $3.4\micron$ and (W2) $4.6\micron$ bands effectively trace photospheric emission from these older stars; W1 is particular sensitive to stellar light, and can often observe luminosities on the order of $\rm L_\star$ out to $z\approx0.5$ \citep{cluver14}. The W1 and W2 bands are optimal tracers of the stellar mass content because they suffer minimally from extinction and can detect the Rayleigh Jeans tail for stars hotter than 2000 K \citep{jarrett13}. We use Eq. (9) from \citet{jarrett13} to derive our stellar masses: $ \rm \log(\frac{M_{*}}{L_{W1}})(\frac{M_\sun}{L_\sun}) = -0.246-2.100(W1-W2)$, where W1 and W2 are given in magnitudes and $\rm L_{W1}$ is the total ``in-band" luminosity, derived by multiplying the spectral luminosity, $\nu L_{3.4}$, by a factor of 22.883; as explained in \citet{jarrett13}, this factor accounts for the difference between the total solar luminosity and the in-band values as measured by WISE. We convert from flux densities (Jy) to magnitudes using the relationship $m_{\rm vega} = -2.5\log_{10}[\frac{f_{\nu}}{f_{\nu0}}]$, with $f_{\nu0}$ given by 306.682 Jy, 170.663 Jy, 29.0448 Jy, and 8.2839 Jy for bands W1, W2, W3, and W4 respectively (see Table 1 in \S IV.4 of the WISE All-Sky Explanatory Supplement\footnote{http://wise2.ipac.caltech.edu/docs/release/allsky/expsup/sec4\_4h.html}). Similarly, we calculate global star formation rates using far-infrared thermal emission (probed by the W4 $22\micron$ band), as a sizable fraction of this emission is powered by young stars. We use Eq. (2) from \citet{jarrett13} to calculate our star formation rates: $\rm SFR_{IR} \;(M_{\sun}\; yr^{-1}) = 7.50 \times 10^{-10} \nu L_{22} \;(L_{\sun})$, where the spectral luminosity has been normalized by $\rm L_{\sun} \;(3.839\; \times 10^{33} \; erg\; s^{-1})$. In all cases, luminosities are derived assuming a distance of $d=4280\; Mpc \times z$, the same distance estimation SDSS recommends for the conversion of SDSS apparent magnitudes to absolute magnitudes.\footnote{http://skyserver.sdss.org/dr12/en/help/cooking/general/getdata5.aspx}

\begin{figure}[H]
\centering
\includegraphics[width=.47\textwidth]{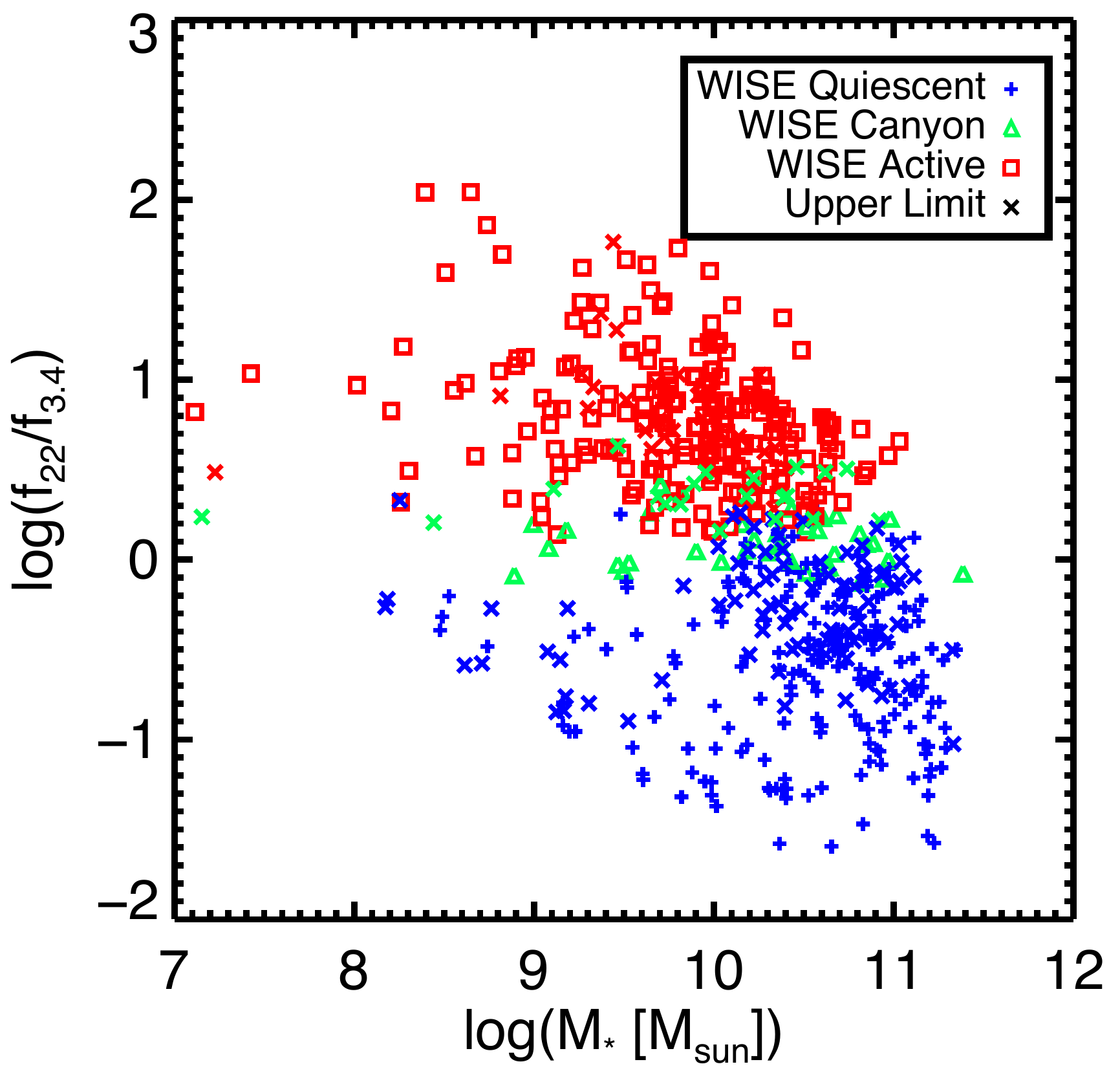}
\includegraphics[width=.465\textwidth]{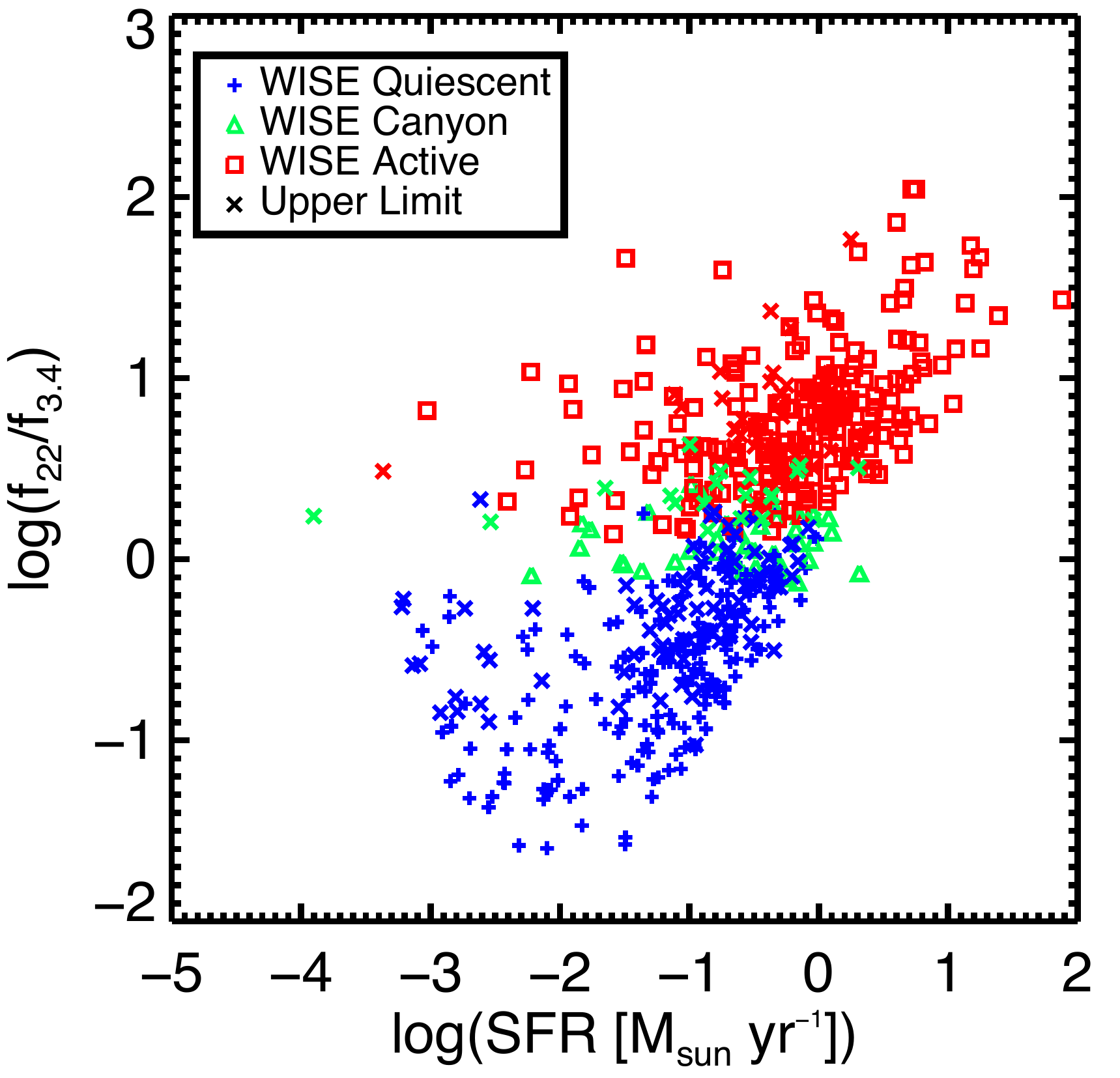}
\caption{We show \textit{Left:} $\log[{\frac{\rm f_{22}}{\rm f_{3.4}}}]$ vs. $\log[\rm M_*]$ and \textit{Right:} $\log[{\frac{\rm f_{22}}{\rm f_{3.4}}}]$ vs. $\log[\rm SFR]$ for our full sample, with the galaxies color-coded by their location in WISE color-color space. Galaxies with an upper limit in either W3 or W4 (see Figure \ref{fig:colorspace_fullsamp}) are color-coded by the same classification but plotted with X's.}
\label{fig:sf_props}
\end{figure}

In Figure \ref{fig:sf_props}, we show $\log[{\frac{\rm f_{22}}{\rm f_{3.4}}}]$ vs. $\log[\rm M_*]$ (effectively sSFR vs. stellar mass) and $\log[{\frac{\rm f_{22}}{\rm f_{3.4}}}]$ vs. $\log[\rm SFR]$ for our full sample. In both cases, the galaxies are color-coded by their location in WISE color-color space.  We note that the WISE quiescent, canyon, and active galaxies from the full sample span approximately the same range in stellar mass ($ \rm \approx 10^8-10^{11.5}\; M_{\sun}$), but the quiescent galaxies (median $\rm M_*$ of $10^{10.59} \; \rm M_\sun$) tend to have systematically higher stellar masses than the WISE canyon galaxies (median $\rm M_*$ of $10^{10.34} \; \rm M_\sun$), while the WISE canyon galaxies have systematically higher stellar masses than the WISE active galaxies (median $\rm M_*$ of $10^{9.92} \; \rm M_\sun$). We also find that the star formation rates of active, canyon, and quiescent galaxies overlap, but that the median of each class is shifted, with the active galaxies predictably having the highest median star formation rate ($\rm 0.65\; M_{\sun}\; yr^{-1}$), followed by the canyon galaxies ($ \rm 0.27\; M_{\sun}\; yr^{-1}$) and the quiescent galaxies ($\rm 0.09\; M_{\sun}\; yr^{-1}$). For the full sample, our star formation rates span $\rm \approx 0.001-10.0 \;M_{\sun}\; yr^{-1}$. 

\subsection{Morphology}
\begin{figure}[H]
\centering
\includegraphics[width=\textwidth]{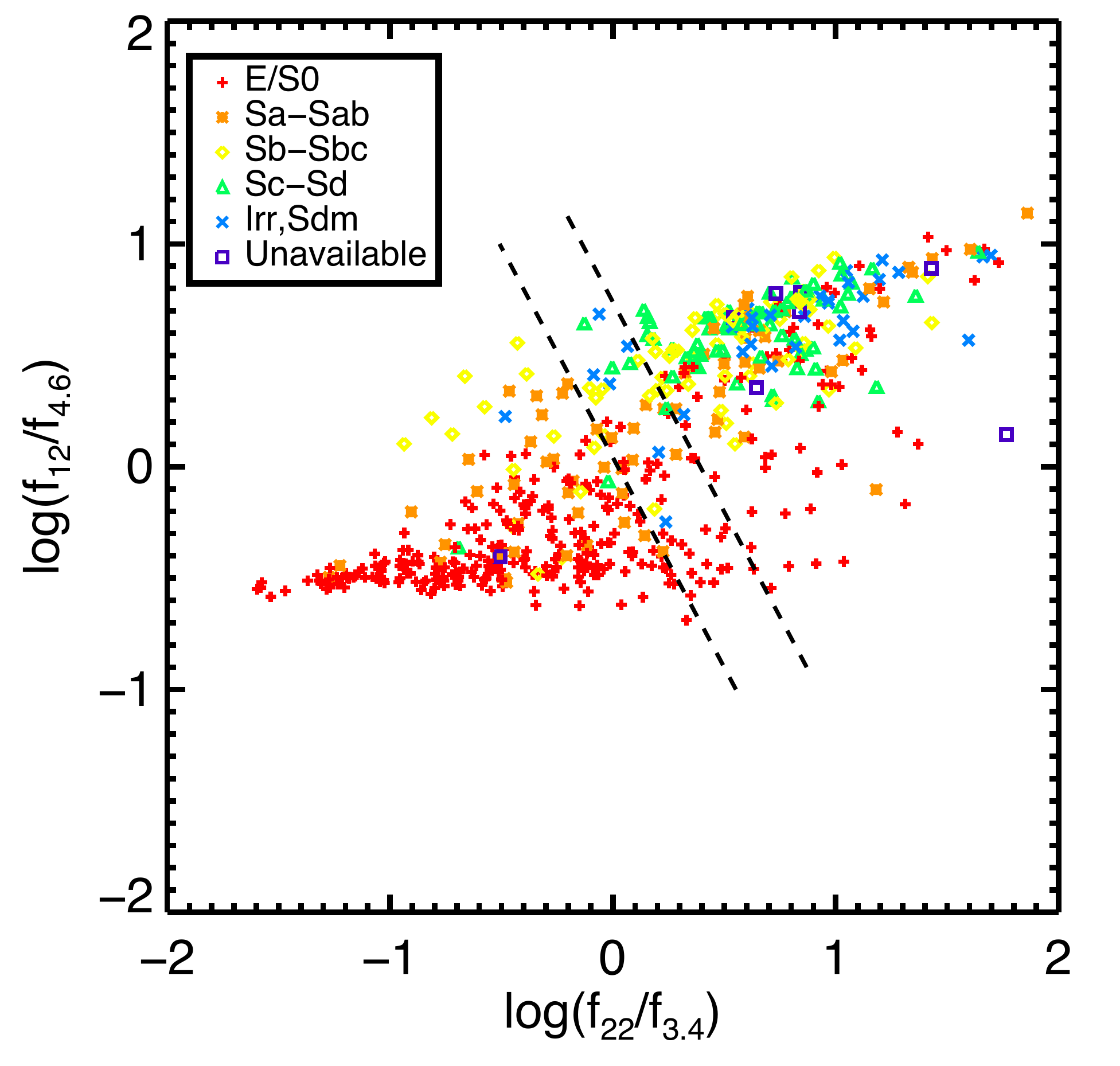}
\caption{Distribution of compact group galaxy morphologies from the full sample in WISE mid-IR colorspace. The boundaries of the canyon are marked with dashed lines. As expected, elliptical and S0 galaxies dominate the quiescent region, while Sc--Sd galaxies dominate the active region. Sa--Sbc types (16 galaxies) and Elliptical/S0 types (13 galaxies) dominate the moderately star forming WISE canyon region, though there is also a small number of Sc-Sd and irregular types (8 galaxies). The lone Sc-Sd galaxy in the quiescent region (IC0107) appears to be an errant misclassification of the galaxy by the HyperLeda morphology compiler, as it morphologically resembles an elliptical galaxy.} 
\label{fig:morphology}
\end{figure}

In order to better understand the physical origins of the WISE canyon, in Figure \ref{fig:morphology} we plot the distribution of galaxy morphologies in WISE colorspace. The morphologies are from the HyperLeda\footnote{http://leda.univ-lyon1.fr/leda/rawcat/a102.html} database \citep{paturel03a,paturel03b}, which accumulates morphologies from available publications and compares and combines them to determine an optimal De Vaucouleurs number for each galaxy. In our classification scheme, an average De Vaucouleurs number between -5.0--0.50 was assigned a morphology of E--S0/a.  Similary, 0.50--2.50=Sa--Sab, 2.50--4.50=Sb--Sbc, 4.50--7.50=Sc--Sd, and 7.50--10.0=Irr,Sdm. In concordance with \citet{alatalo14}, we see a significant bifurcation between spirals and bulge-dominated galaxies in this distinct region of WISE color-color space: the quiescent region is dominated by elliptical and S0 galaxies, while the majority of Sc--Sd spiral and irregular galaxies occupy the active region. Excluding those with upper limits, the canyon galaxies between these two populations exhibit a diverse range of morphologies, but is dominated in almost equal numbers by Sa-Sbc types (16 galaxies) and Elliptical/S0 types (13 galaxies). The rest of the canyon galaxies with reliable photometry are a mix of Sc-Sd type (4 galaxies) and irregulars (4 galaxies). A large fraction of the upper limit canyon galaxies (17/20 galaxies) are Elliptical/S0s, and would produce a canyon that is overwhelmingly early type if included in this analysis. The abundance of Sa-Sbc type galaxies in the higher signal-to-noise canyon region is not surprising, as these galaxies tend to have brighter mid-IR-blue bulges than Sc-Sd type galaxies \citep{jarrett00}. When combined with their actively to moderately star-forming mid-IR-green and red disks, this combination would be expected to produce green mid-IR colors indicative of moderate star formation.

In Figure \ref{fig:thumbnails}, we show optical, five-color SDSS thumbnails\footnote{http://skyserver.sdss.org/dr12/en/tools/chart/listinfo.aspx} alongside their WISE three-color thumbnails\footnote{https://irsa.ipac.caltech.edu/applications/wise/} for all the WISE canyon galaxies with reliable photometry ($SNR > 2$ in all four bands) in the SDSS DR12 footprint (31 of 37 total canyon galaxies). Of the 31 reliable canyon galaxies within the SDSS footprint, 11 show visible signs of gravitational interaction with a neighboring companion in their WISE thumbnail. Moreover, the combination of mid-IR blue bulges and mid-IR red star-forming disks is apparent in a number of the WISE thumbnails (for example, HCG22c, HCG45a, HCG69a, HCG71a, HCG88a, NGC0070, NGC4274). Interestingly, several galaxies show the opposite color combination (mid-IR red nucleus and mid-IR blue outskirts), including HCG37b, NGC4117, and NGC4314. 

\doublespace
\clearpage

\begin{center}
\LTcapwidth=\textwidth
\begin{longtable}{ccc}
\caption*{WISE-SDSS Thumbnail Comparison\label{thumbnails}}\\
\tiny HGC01b & \tiny HCG05a & \tiny HCG06b\\
\includegraphics[width=2cm,height=2cm]{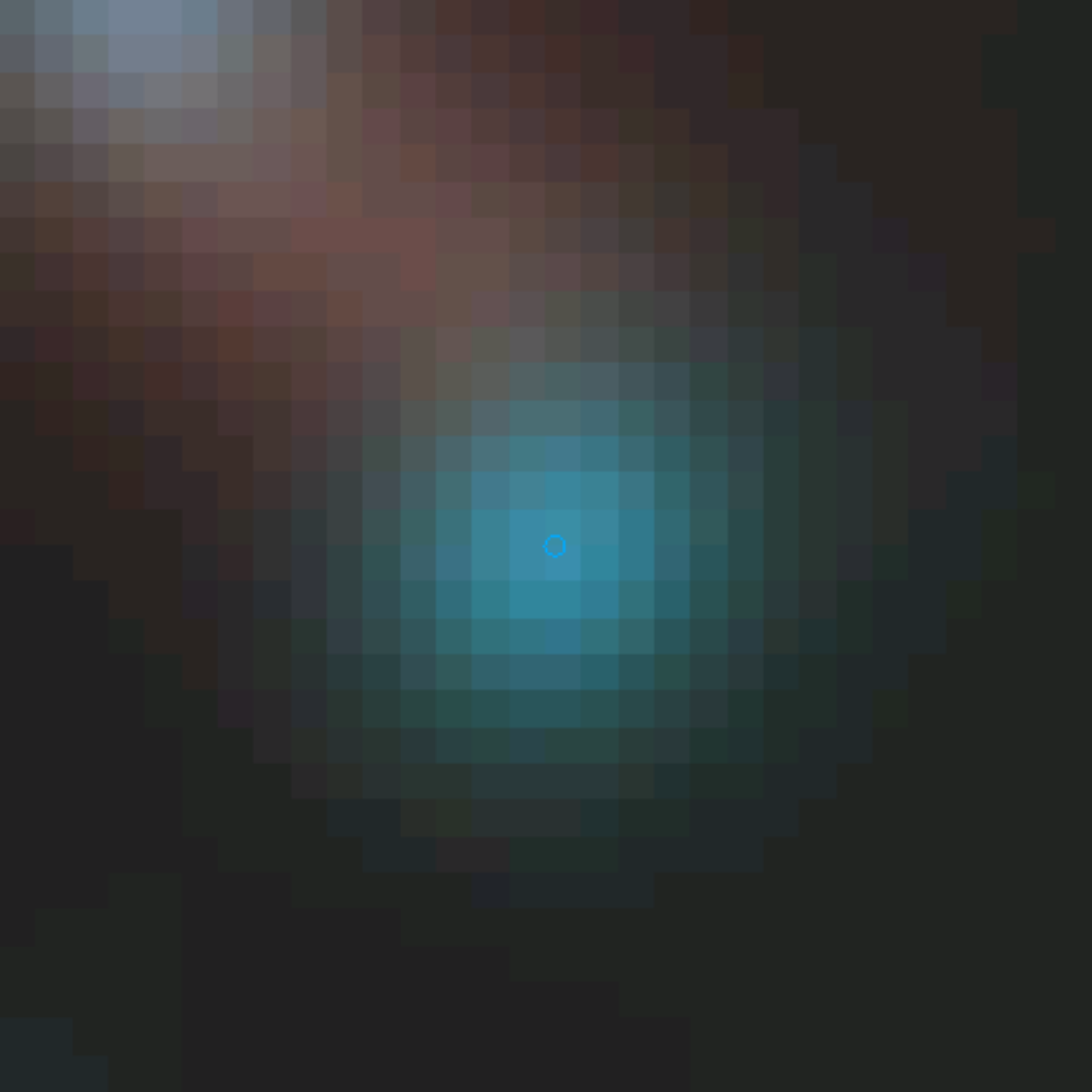}\hspace*{-.1em}
\includegraphics[width=2cm,height=2cm]{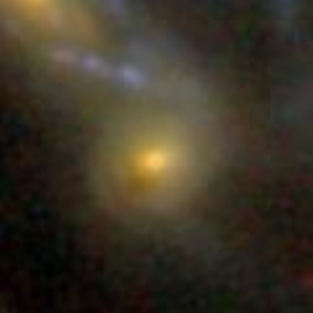}\hspace*{-.1em}&
\includegraphics[width=2cm,height=2cm]{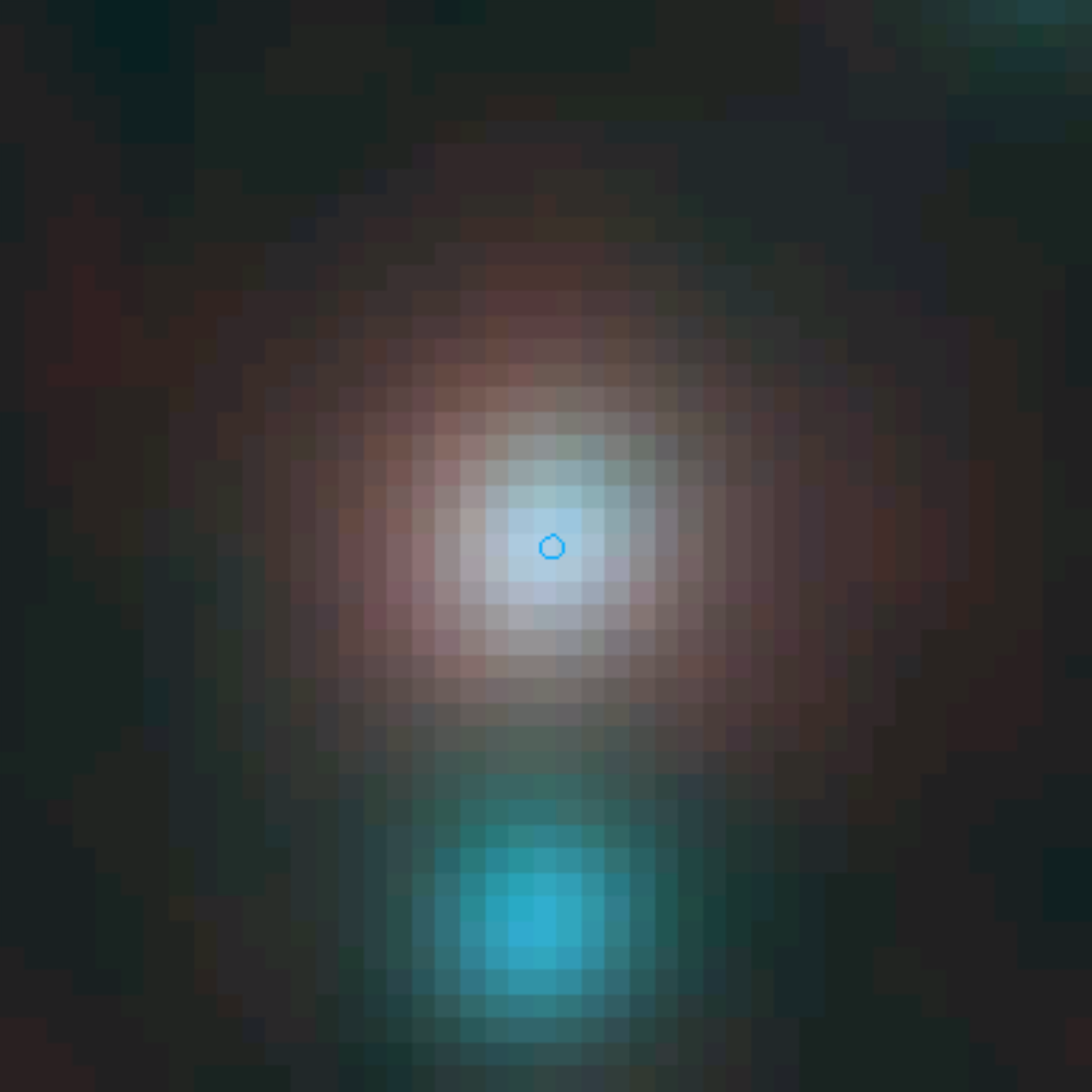}\hspace*{-.1em}
\includegraphics[width=2cm,height=2cm]{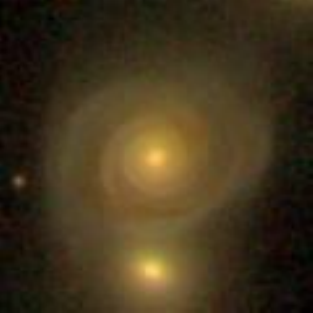}\hspace*{-.1em}&
\includegraphics[width=2cm,height=2cm]{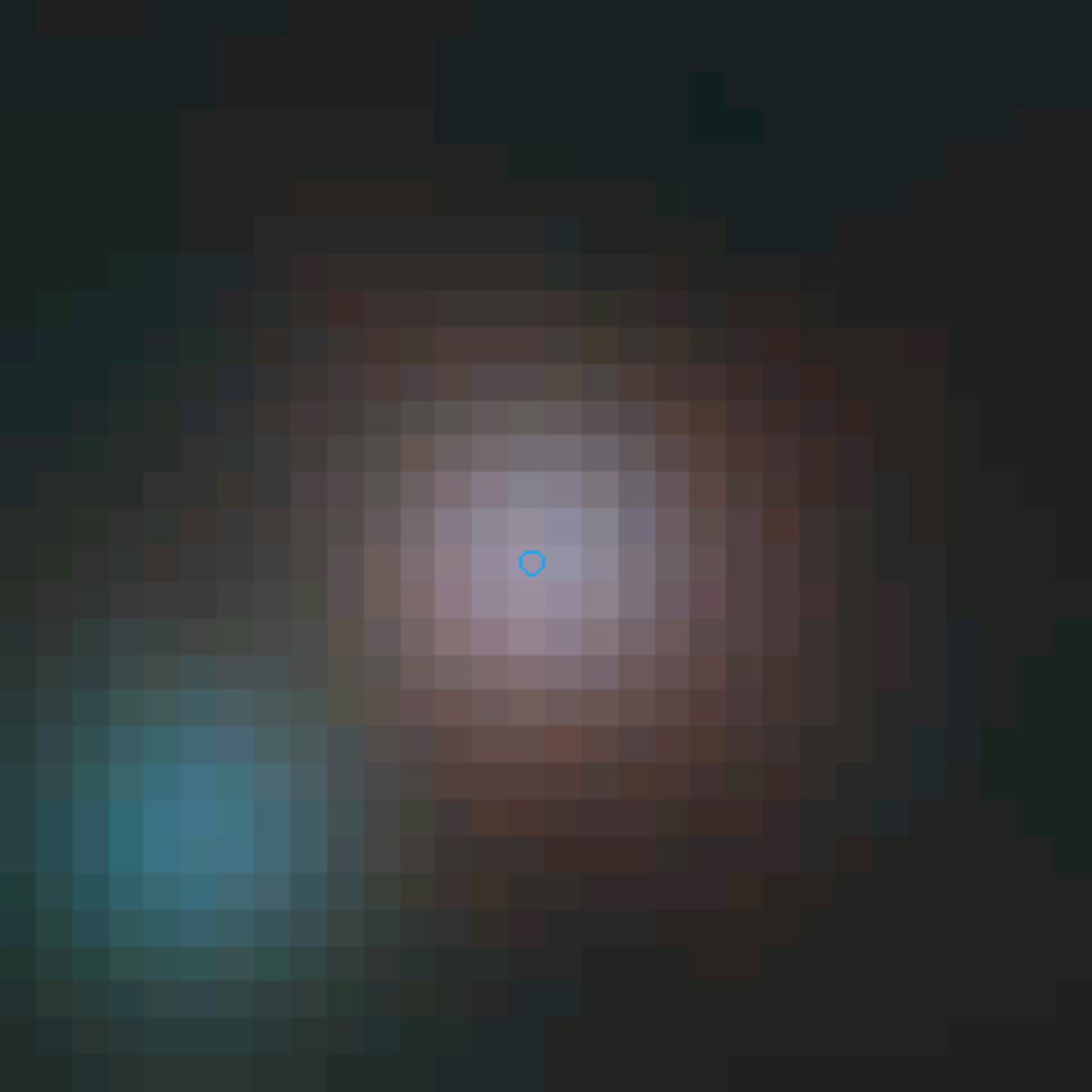}\hspace*{-.1em}
\includegraphics[width=2cm,height=2cm]{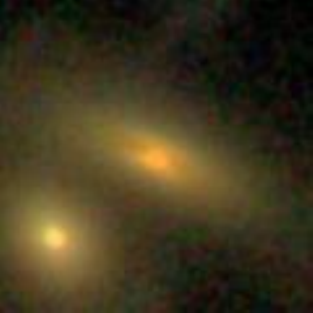}\hspace*{-.1em}\\[-.2in]
\tiny HCG22c & \tiny HCG37b & \tiny HCG44a\\
\includegraphics[width=2cm,height=2cm]{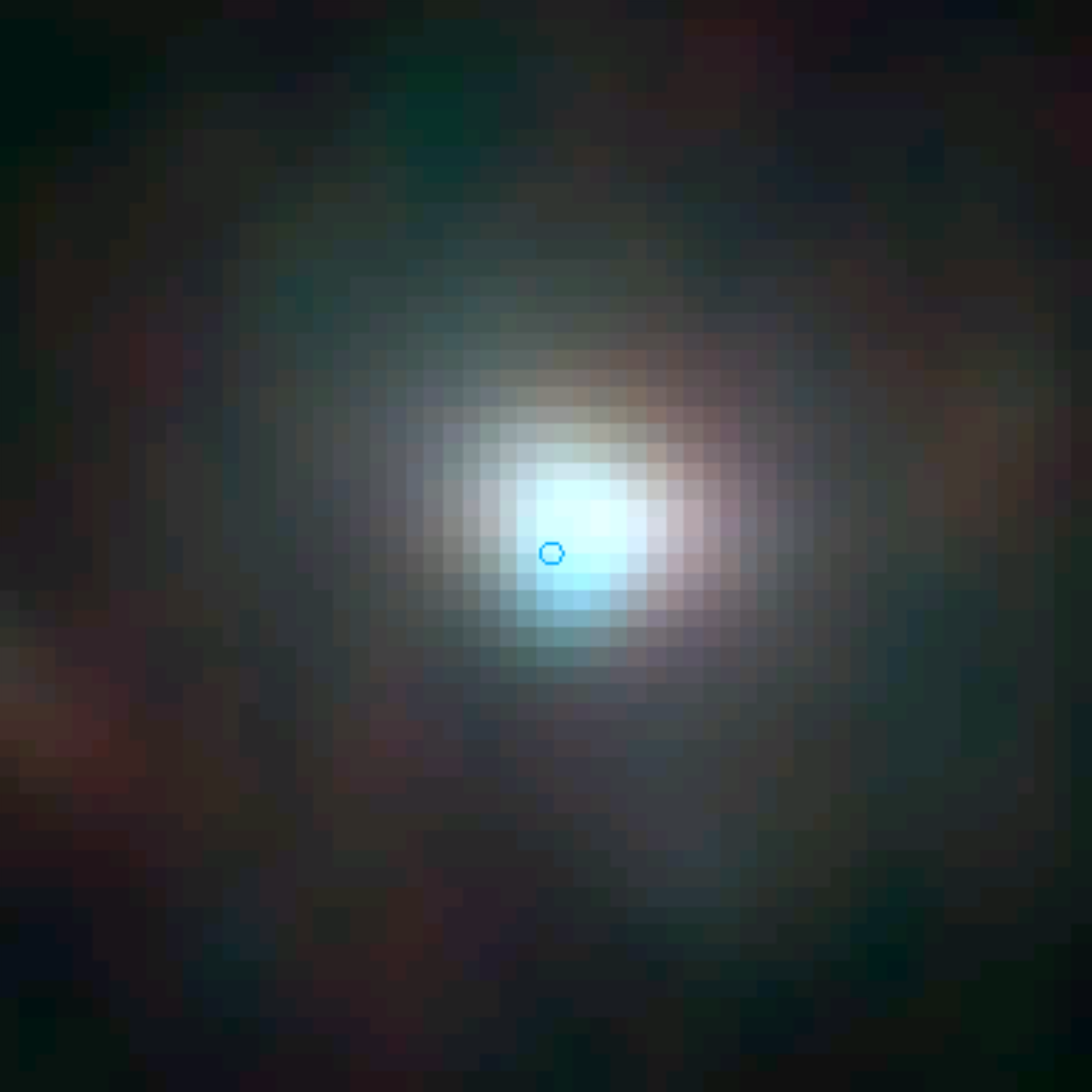}\hspace*{-.1em}
\includegraphics[width=2cm,height=2cm]{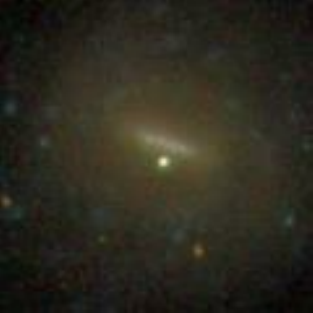}\hspace*{-.1em}&
\includegraphics[width=2cm,height=2cm]{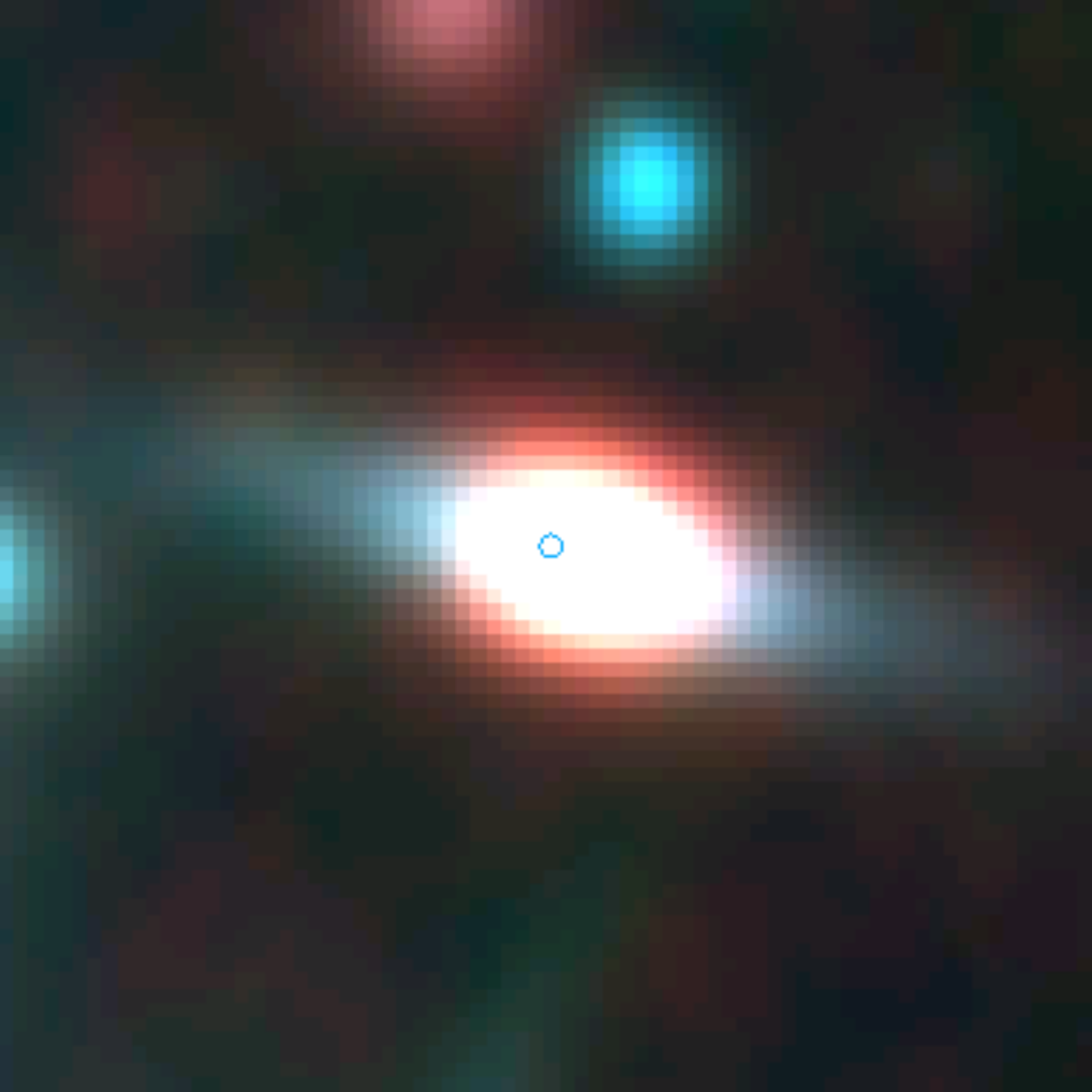}\hspace*{-.1em}
\includegraphics[width=2cm,height=2cm]{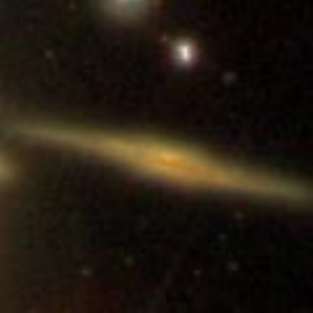}\hspace*{-.1em}&
\includegraphics[width=2cm,height=2cm]{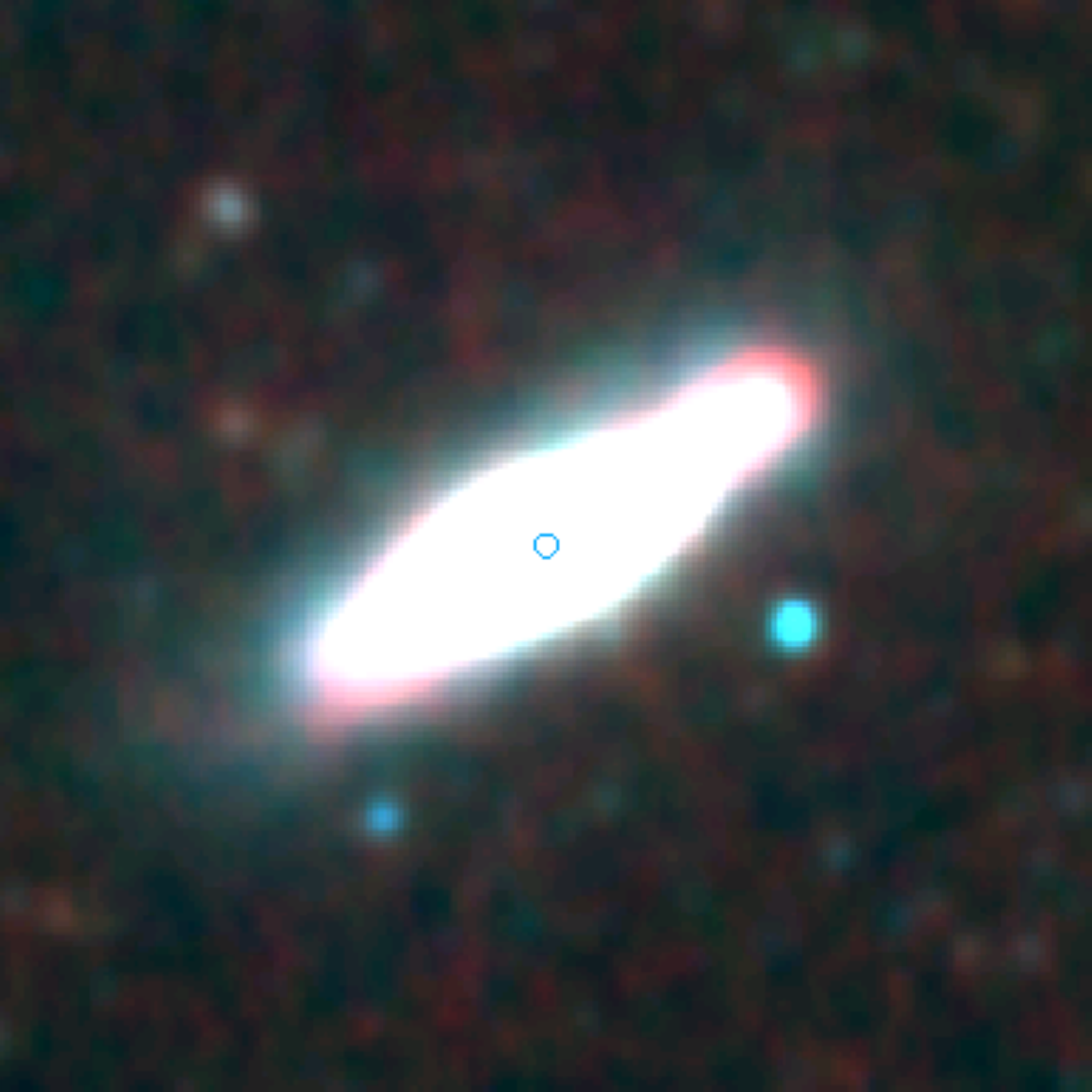}\hspace*{-.1em}
\includegraphics[width=2cm,height=2cm]{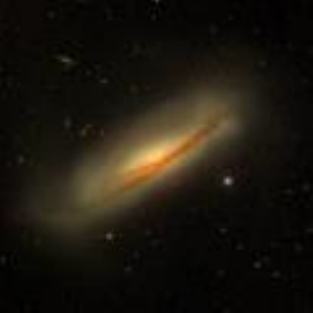}\hspace*{-.1em}\\[-.2in]
\tiny HCG45a & \tiny HCG45b & \tiny HCG46b\\
\includegraphics[width=2cm,height=2cm]{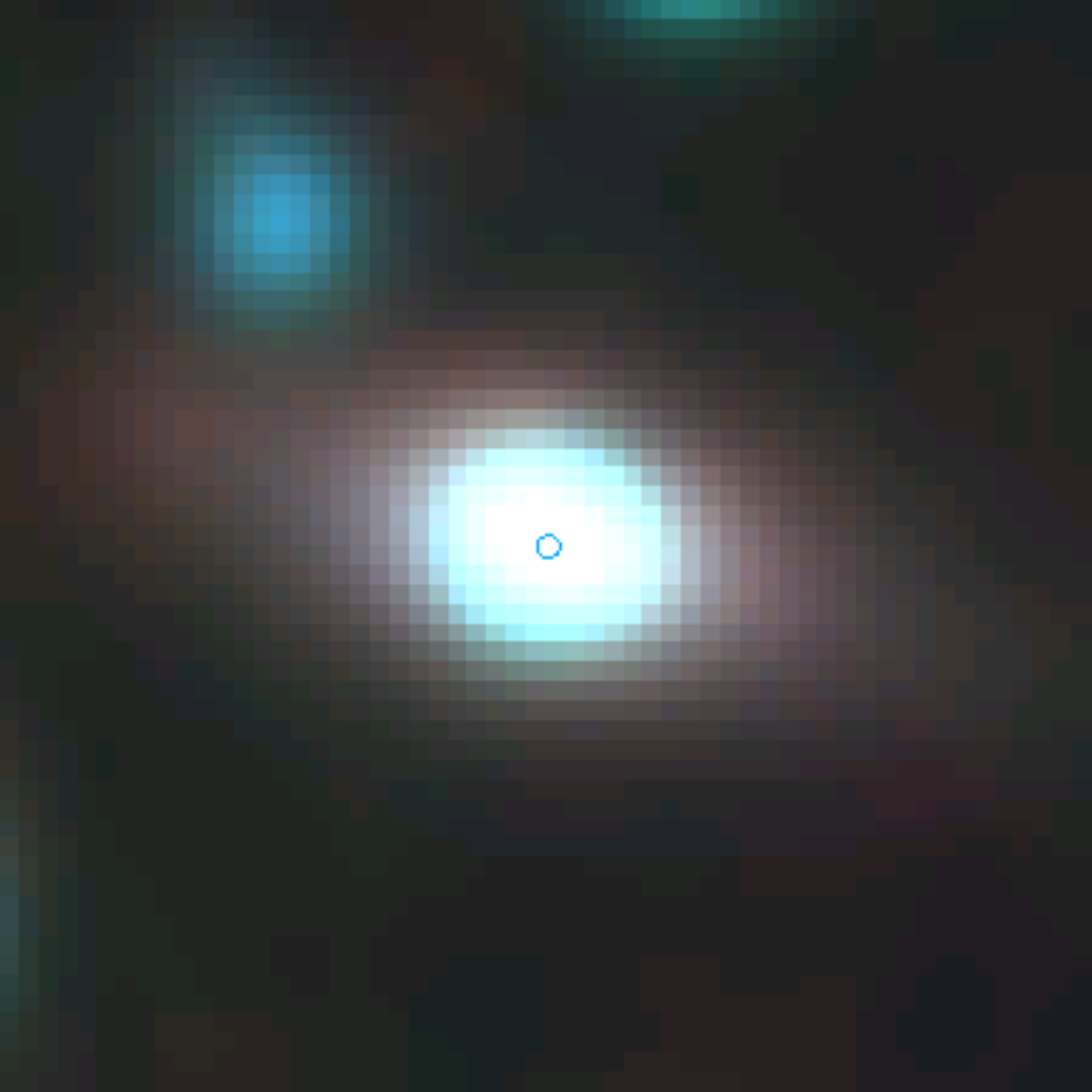}\hspace*{-.1em}
\includegraphics[width=2cm,height=2cm]{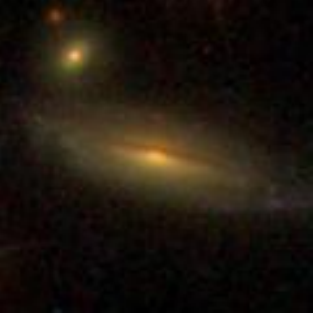}\hspace*{-.1em}&
\includegraphics[width=2cm,height=2cm]{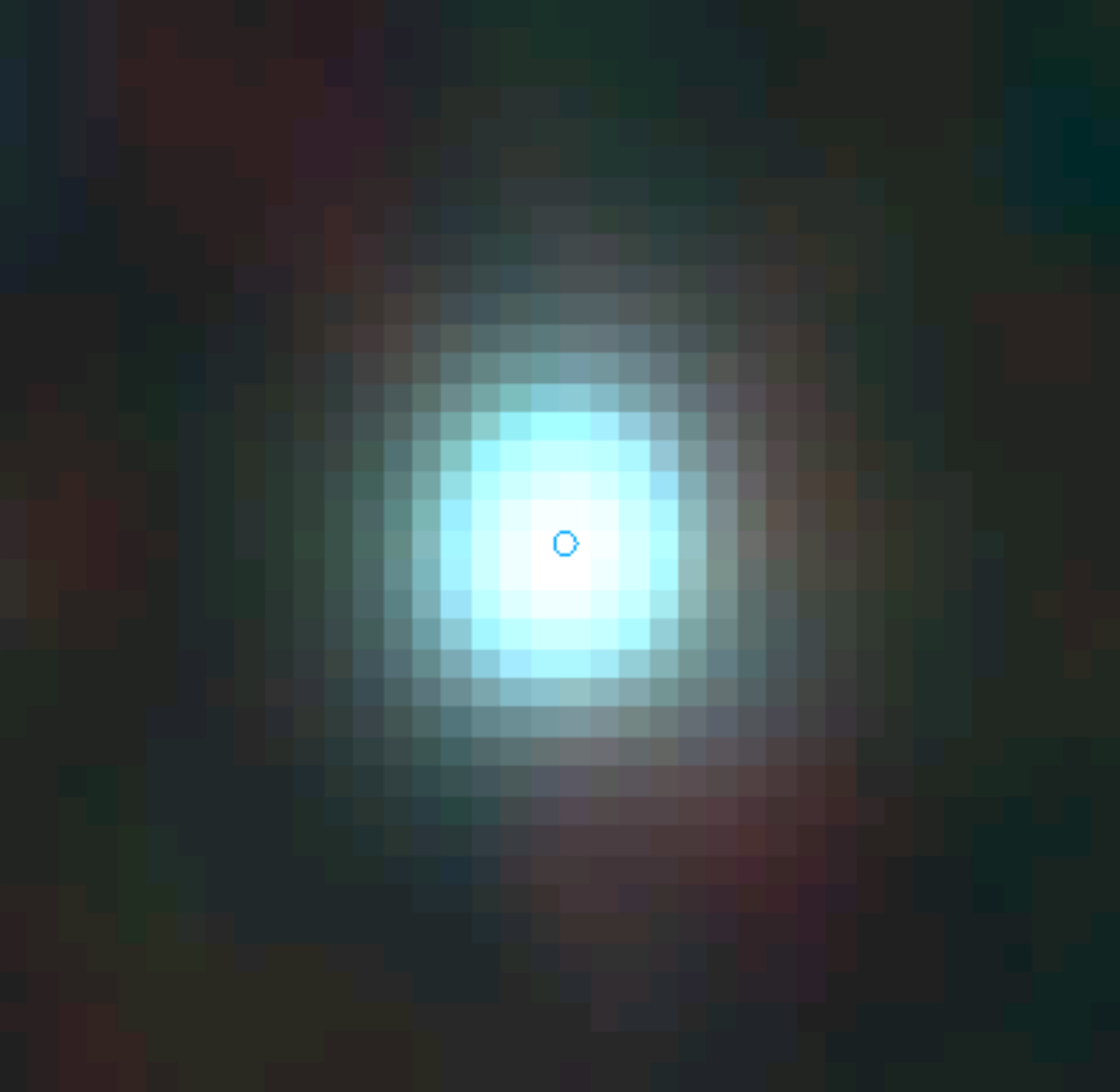}\hspace*{-.1em}
\includegraphics[width=2cm,height=2cm]{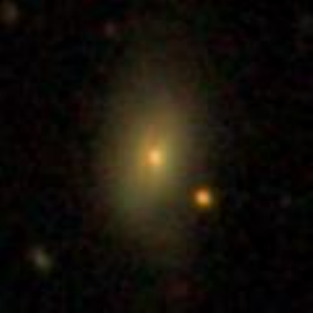}\hspace*{-.1em}&
\includegraphics[width=2cm,height=2cm]{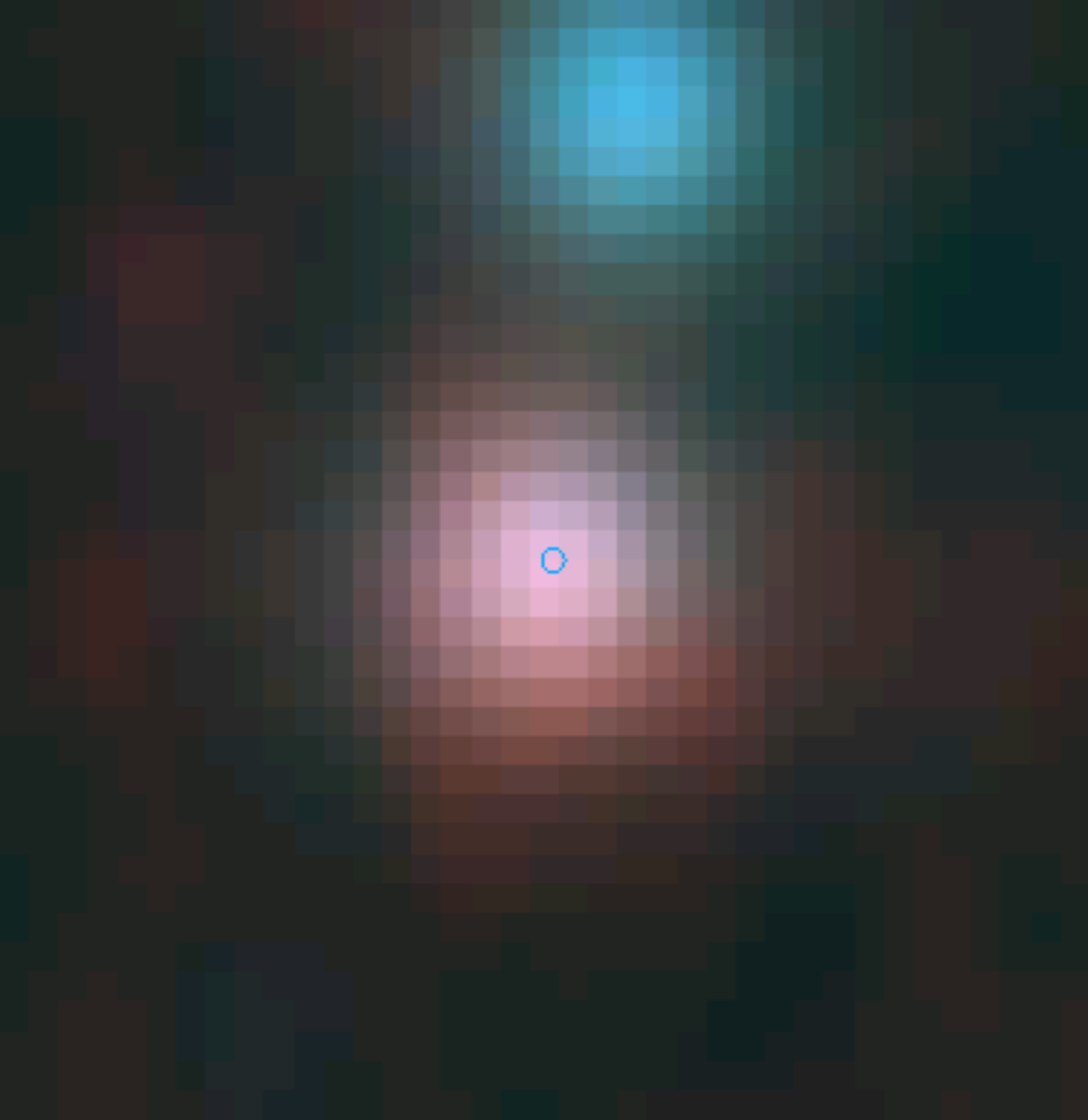}\hspace*{-.1em}
\includegraphics[width=2cm,height=2cm]{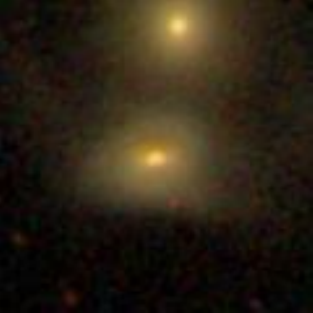}\hspace*{-.1em}\\[-.2in]
\tiny HCG51b & \tiny HCG57a & \tiny HCG59b\\
\includegraphics[width=2cm,height=2cm]{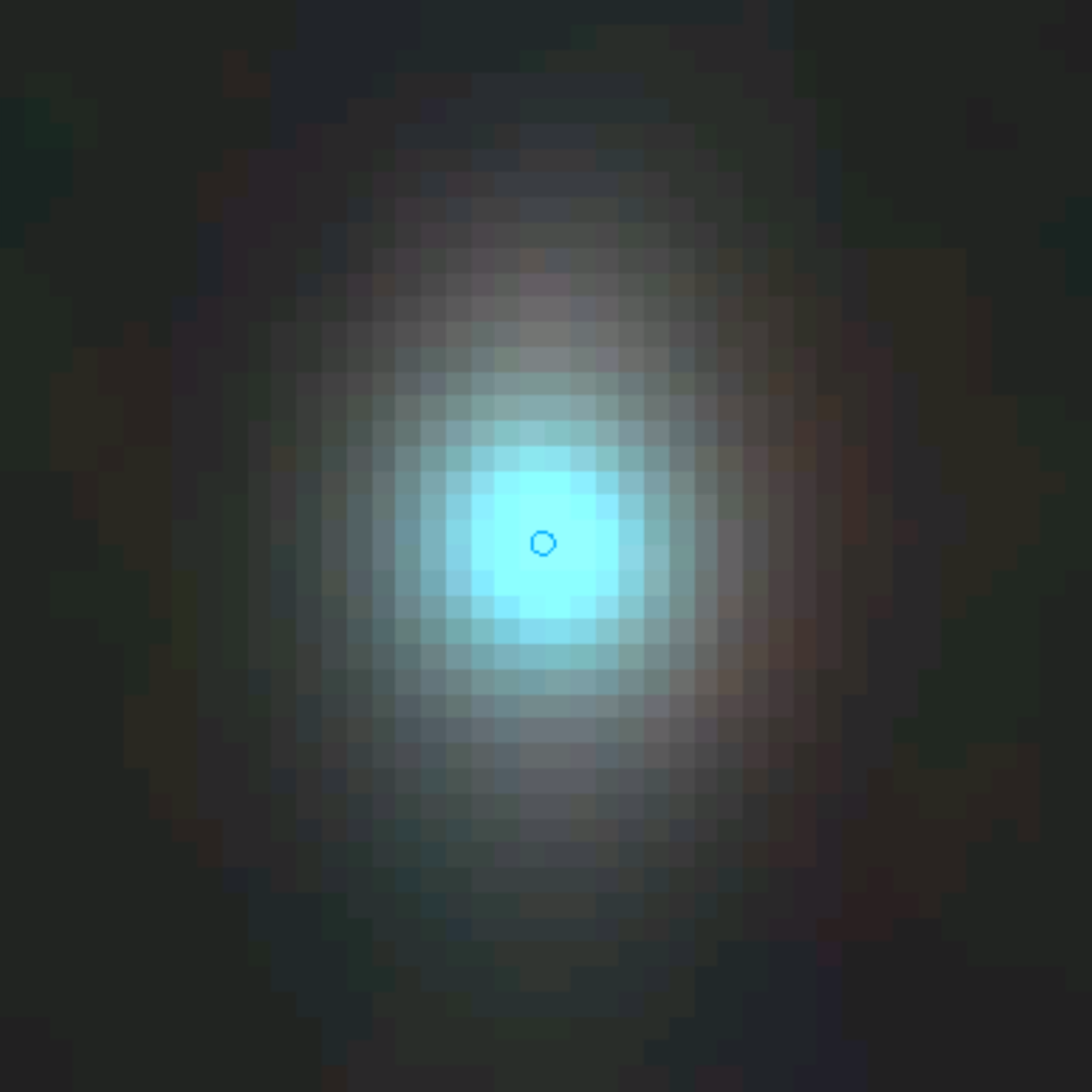}\hspace*{-.1em}
\includegraphics[width=2cm,height=2cm]{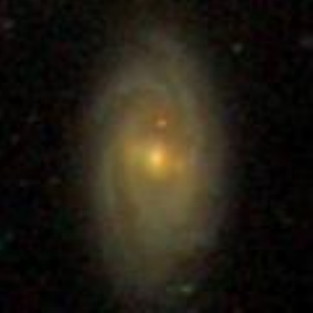}\hspace*{-.1em}&
\includegraphics[width=2cm,height=2cm]{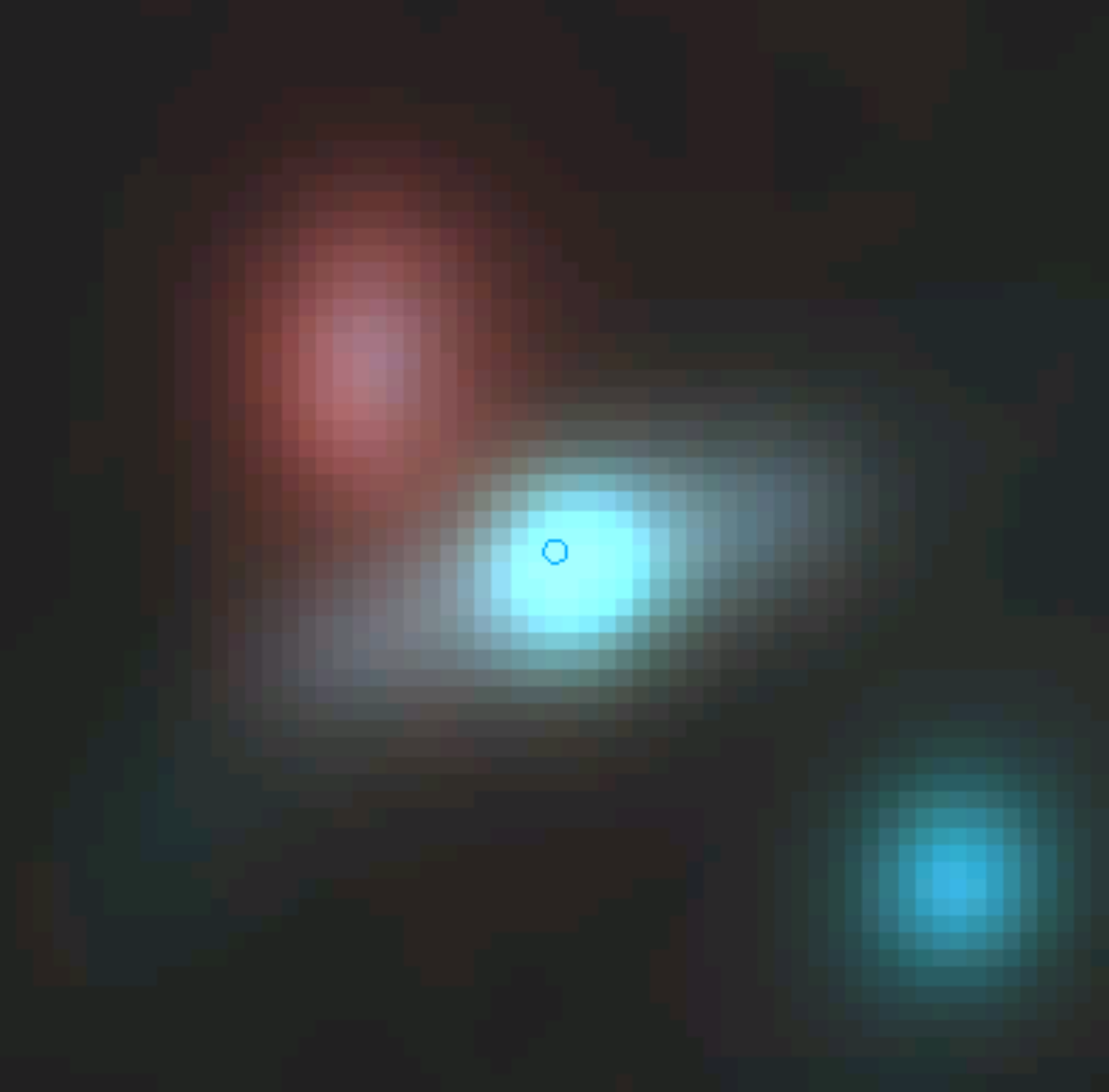}\hspace*{-.1em}
\includegraphics[width=2cm,height=2cm]{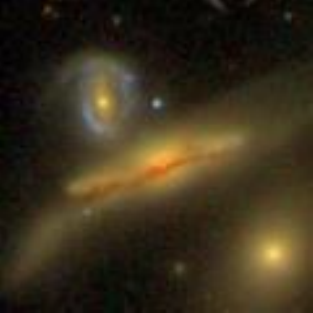}\hspace*{-.1em}&
\includegraphics[width=2cm,height=2cm]{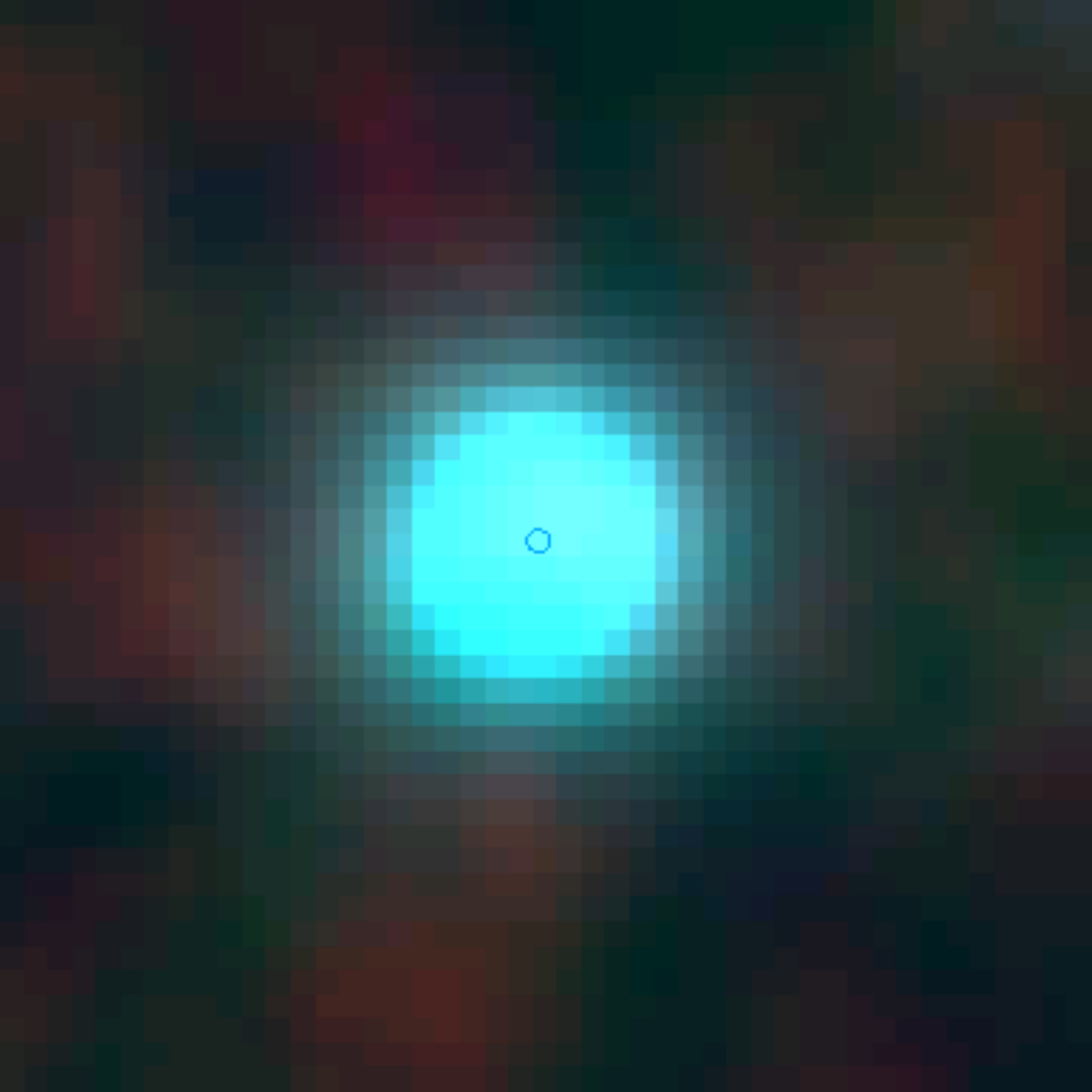}\hspace*{-.1em}
\includegraphics[width=2cm,height=2cm]{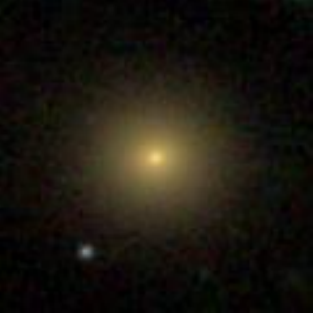}\hspace*{-.1em}\\[-.2in]
\tiny HCG60b & \tiny HCG69a & \tiny HCG70c \\
\includegraphics[width=2cm,height=2cm]{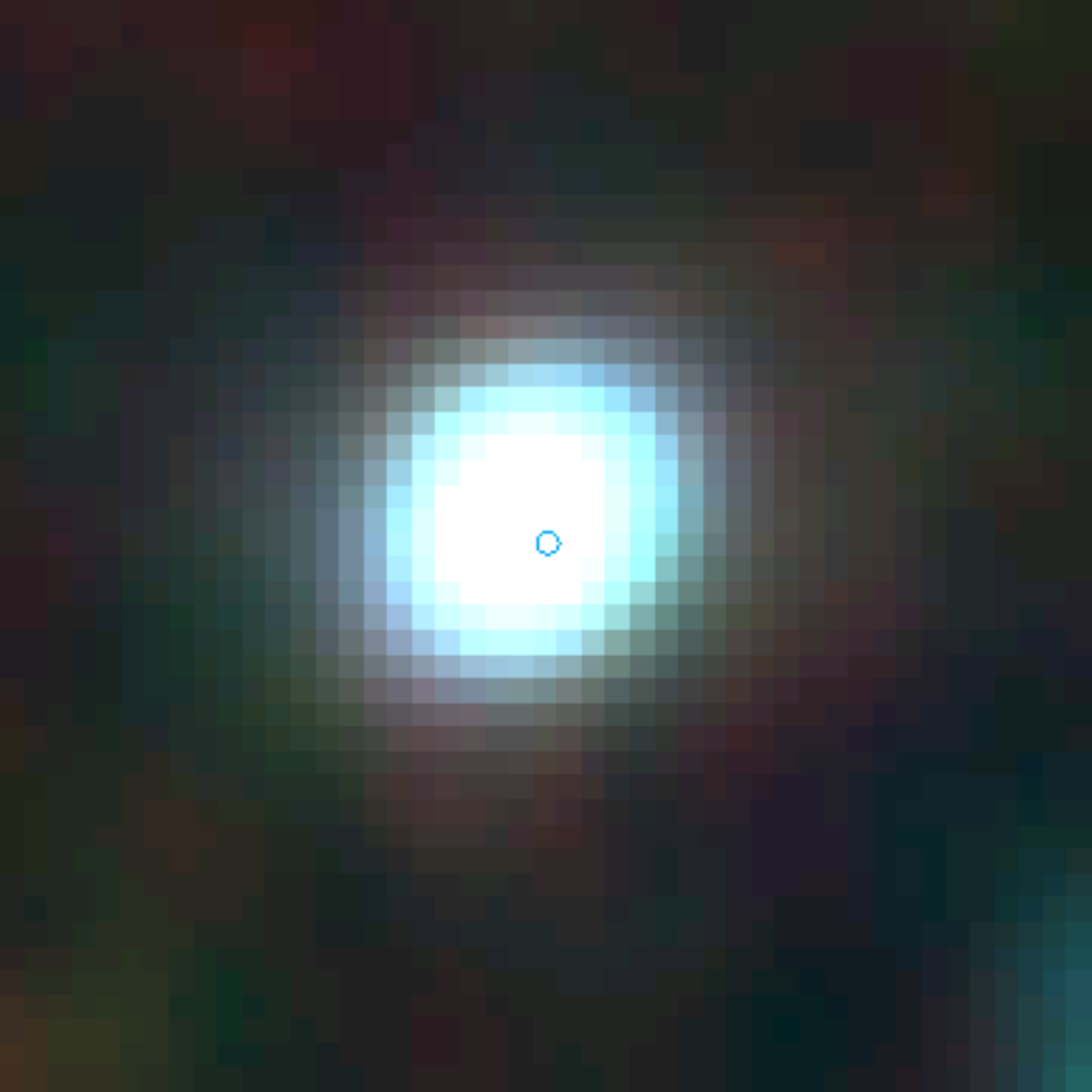}\hspace*{-.1em}
\includegraphics[width=2cm,height=2cm]{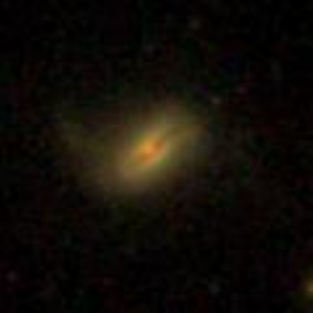}\hspace*{-.1em}&
\includegraphics[width=2cm,height=2cm]{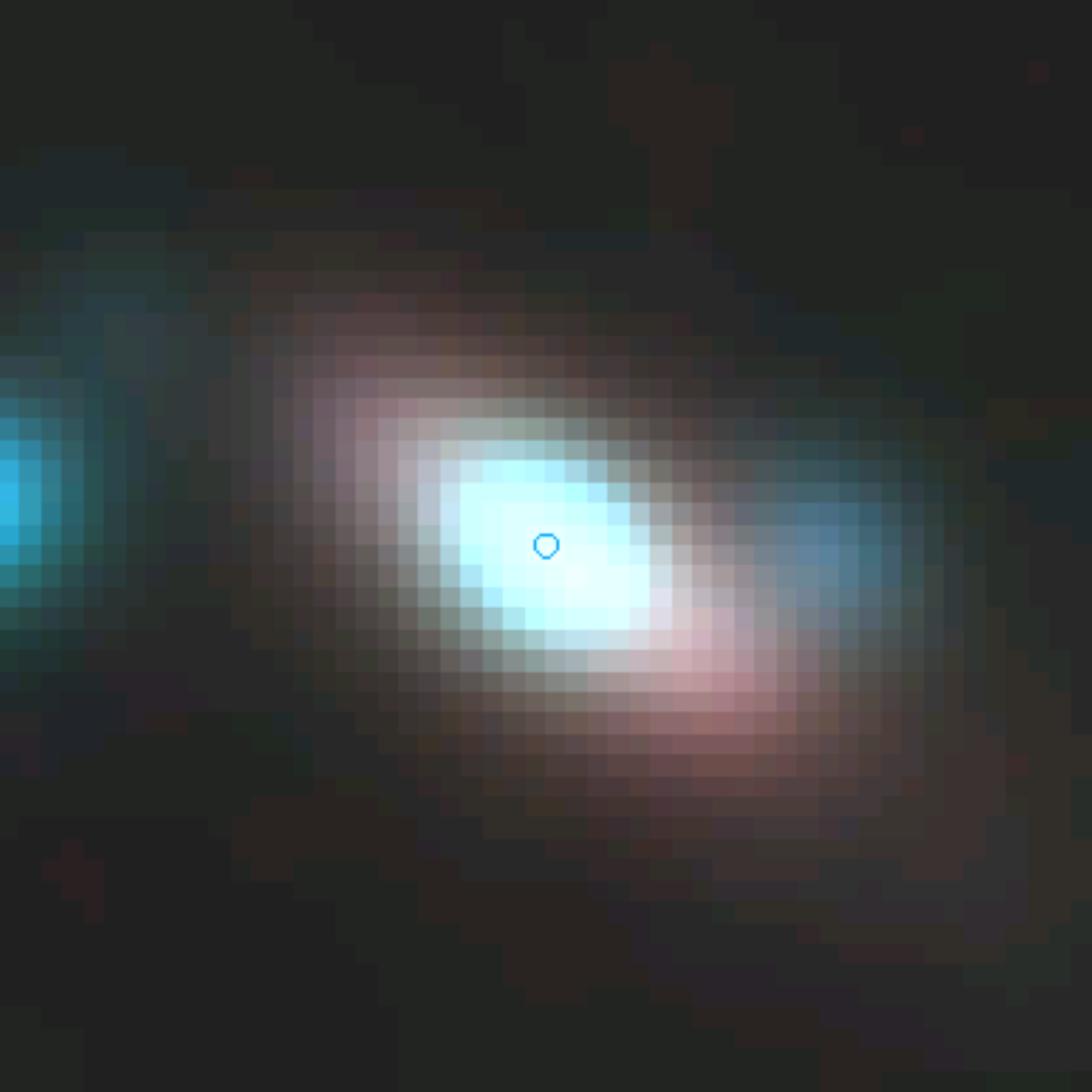}\hspace*{-.1em}
\includegraphics[width=2cm,height=2cm]{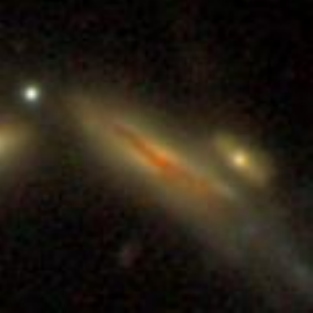}\hspace*{-.1em}&
\includegraphics[width=2cm,height=2cm]{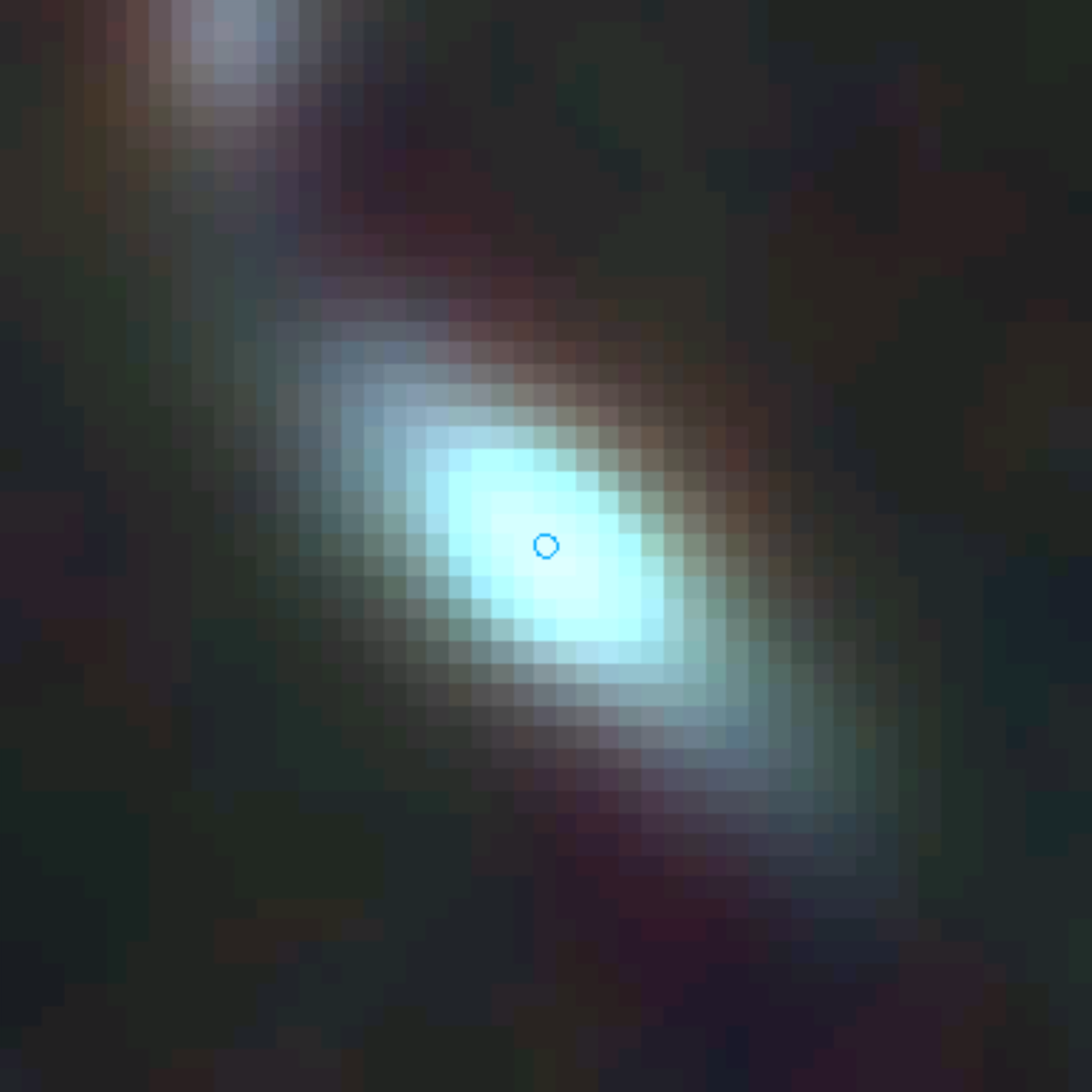}\hspace*{-.1em}
\includegraphics[width=2cm,height=2cm]{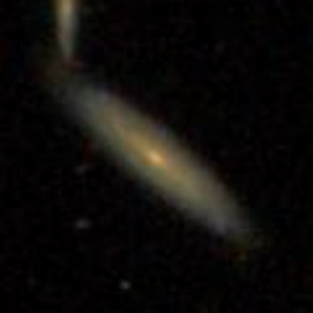}\hspace*{-.1em}\\[-.2in]
\tiny HCG71a & \tiny HCG75c & \tiny HCG81c\\
\includegraphics[width=2cm,height=2cm]{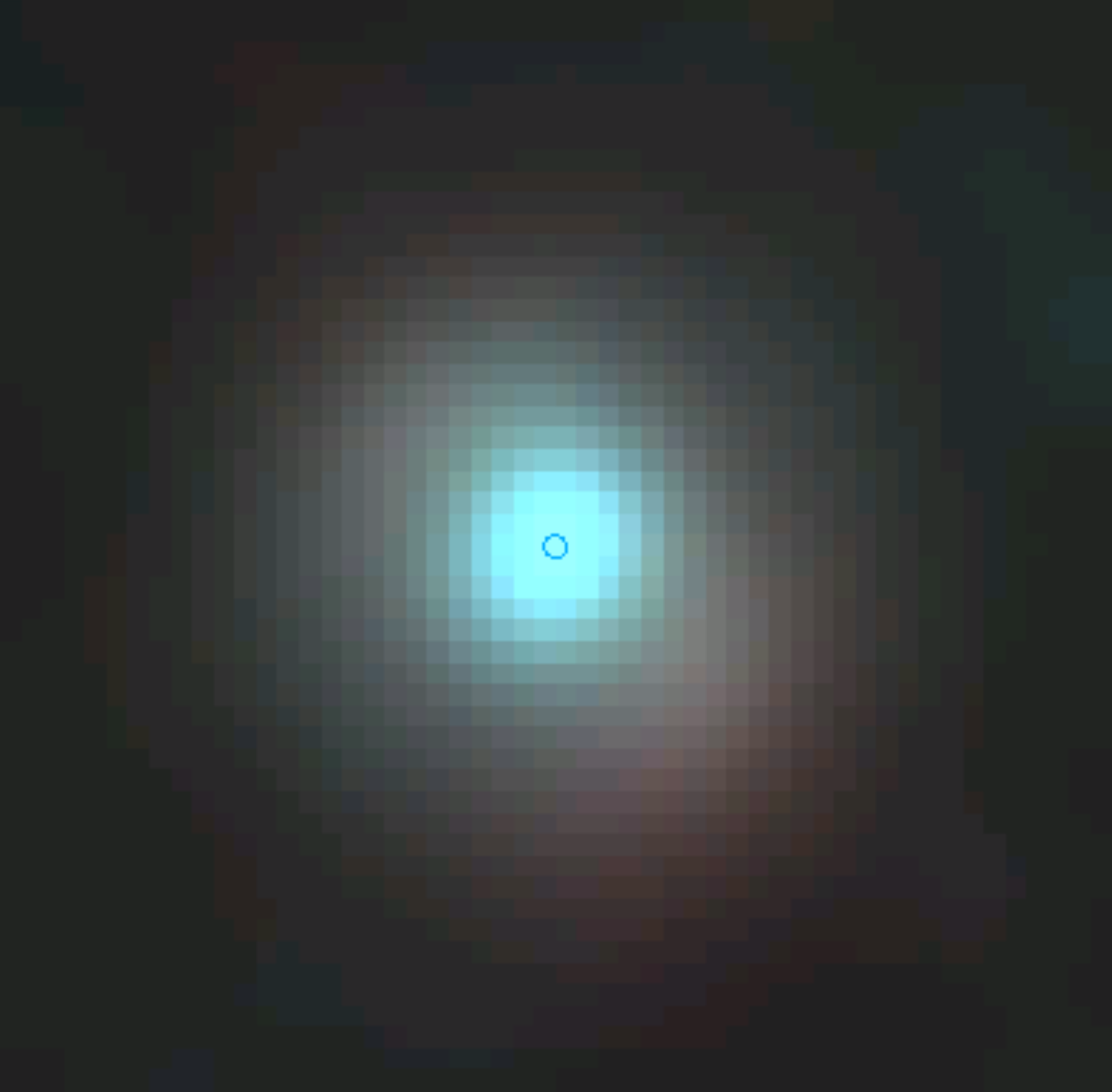}\hspace*{-.1em}
\includegraphics[width=2cm,height=2cm]{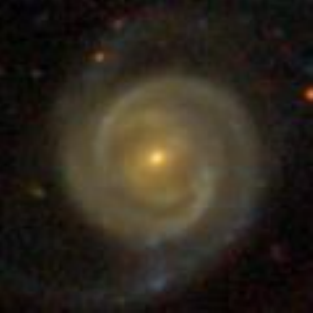}\hspace*{-.1em}&
\includegraphics[width=2cm,height=2cm]{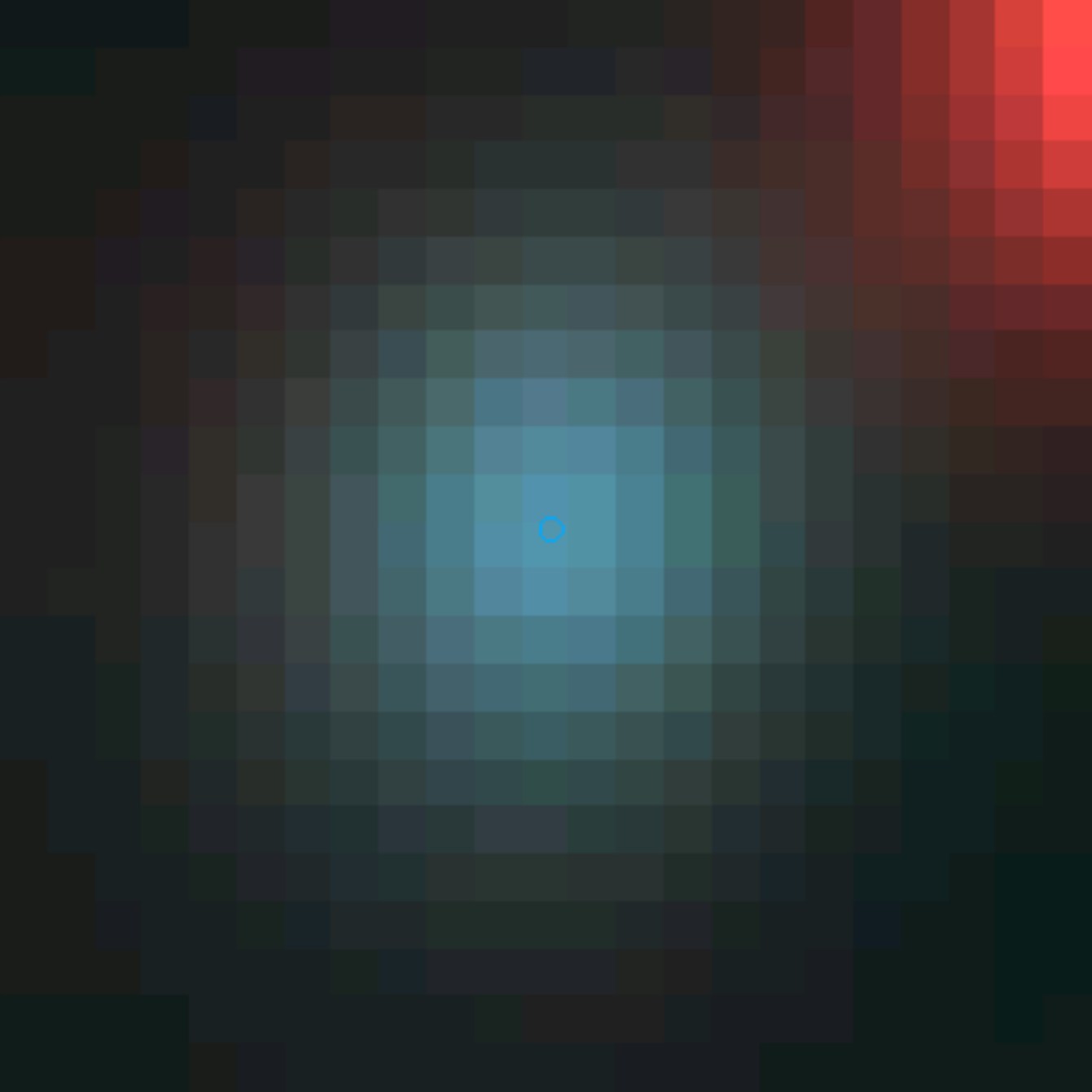}\hspace*{-.1em}
\includegraphics[width=2cm,height=2cm]{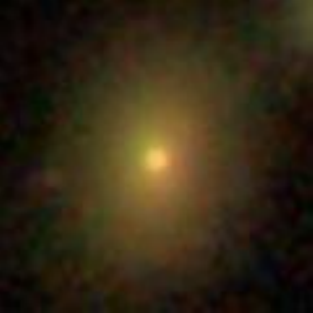}\hspace*{-.1em}&
\includegraphics[width=2cm,height=2cm]{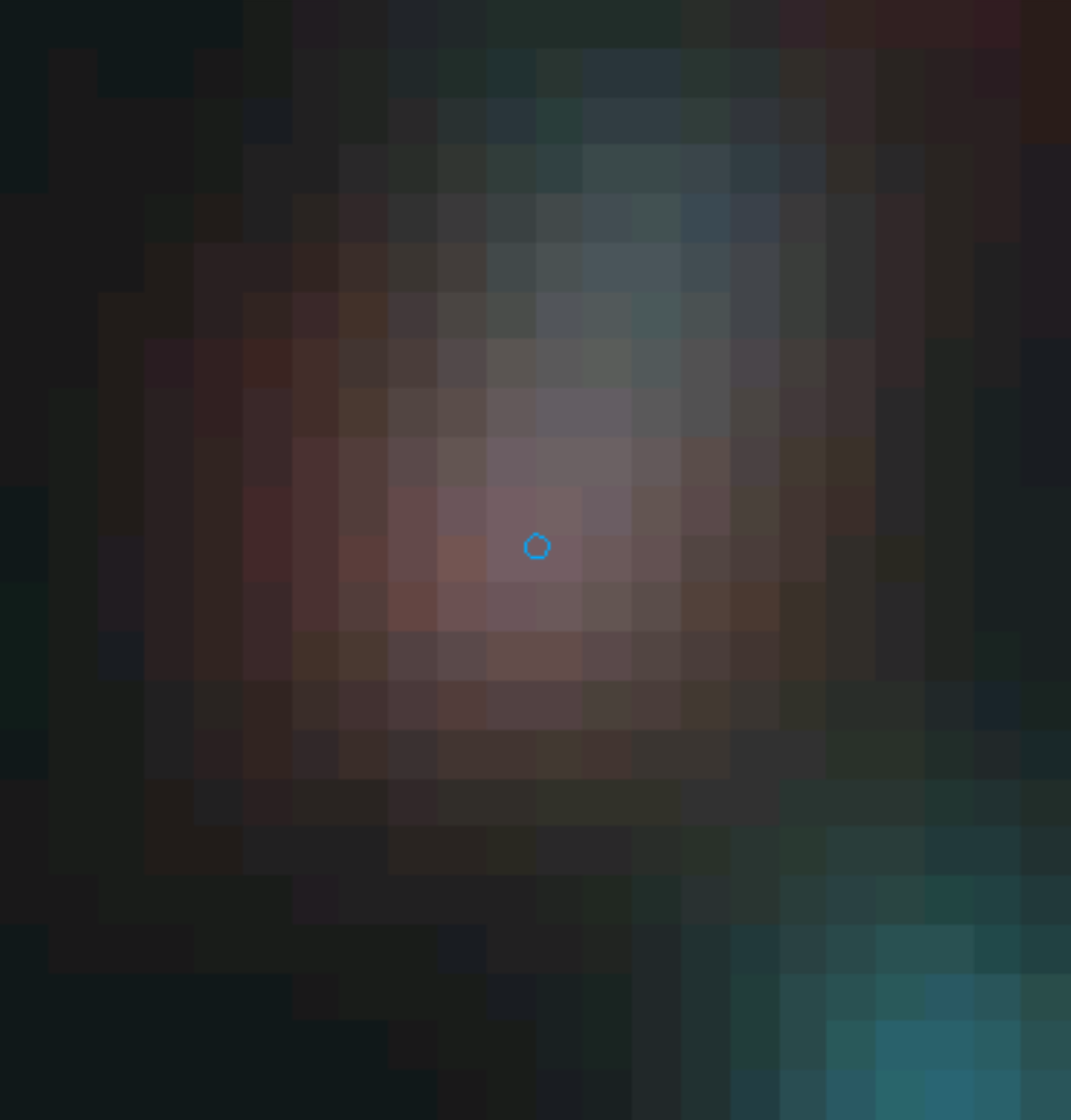}\hspace*{-.1em}
\includegraphics[width=2cm,height=2cm]{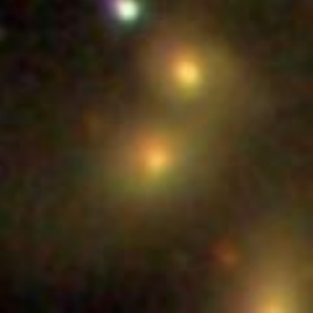}\hspace*{-.1em}\\[-.2in]
\tiny HCG88a & \tiny HCG97b &  \tiny HCG99d\\
\includegraphics[width=2cm,height=2cm]{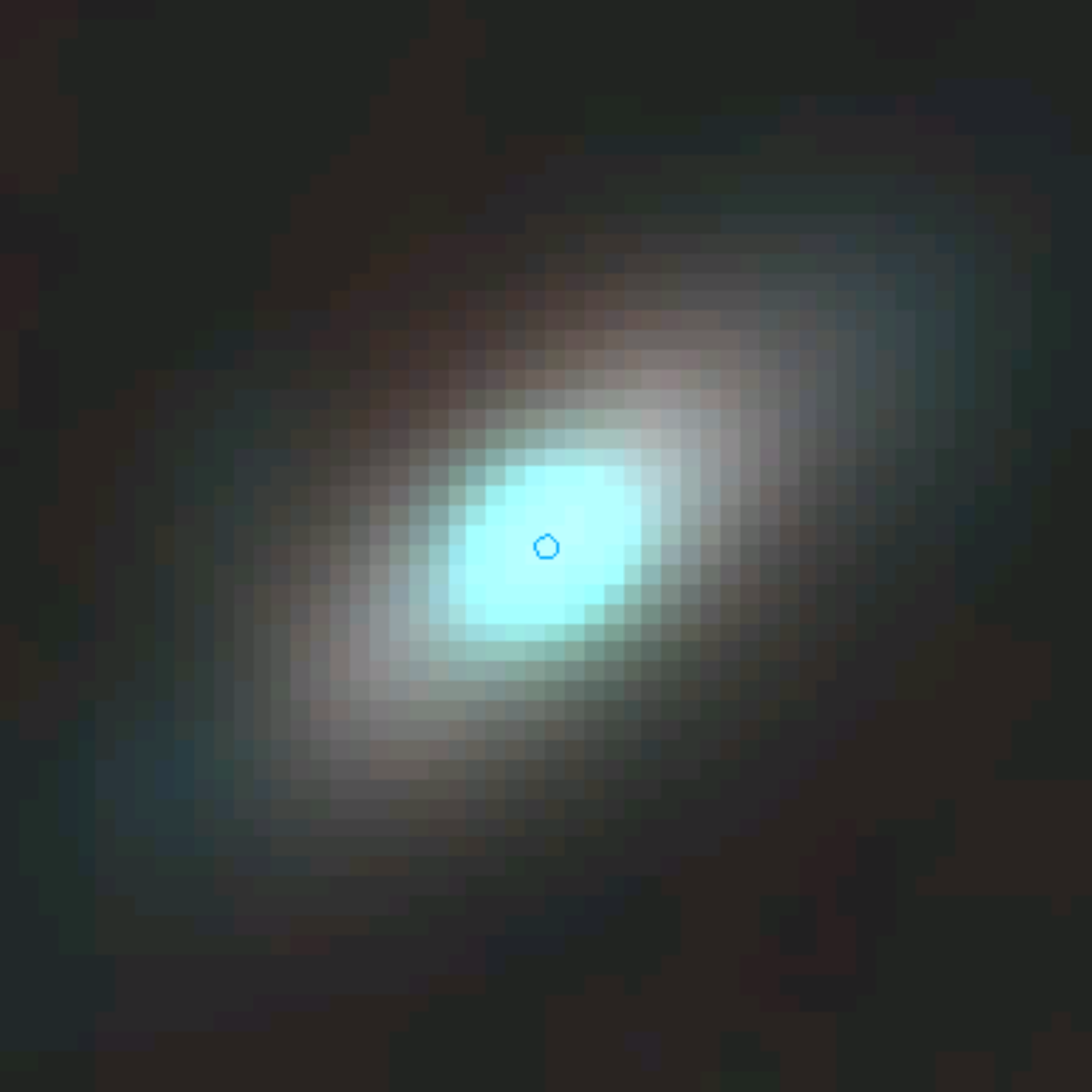}\hspace*{-.1em}
\includegraphics[width=2cm,height=2cm]{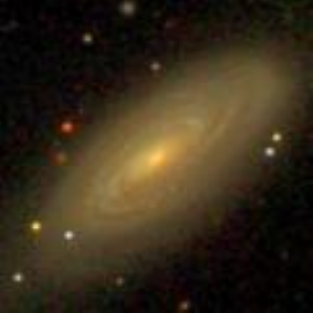}\hspace*{-.1em}&
\includegraphics[width=2cm,height=2cm]{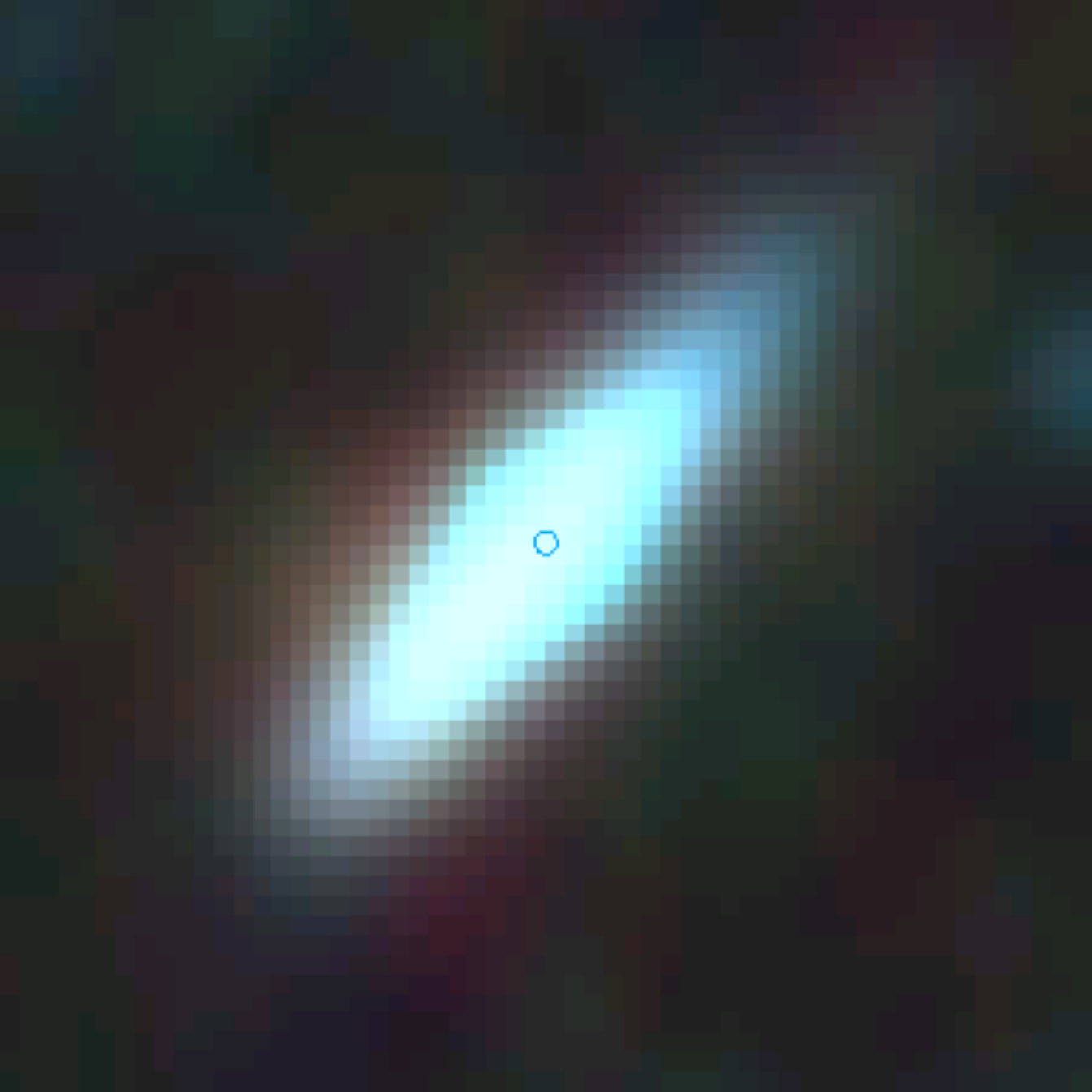}\hspace*{-.1em}
\includegraphics[width=2cm,height=2cm]{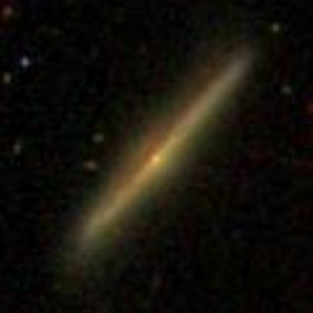}\hspace*{-.1em}&
\includegraphics[width=2cm,height=2cm]{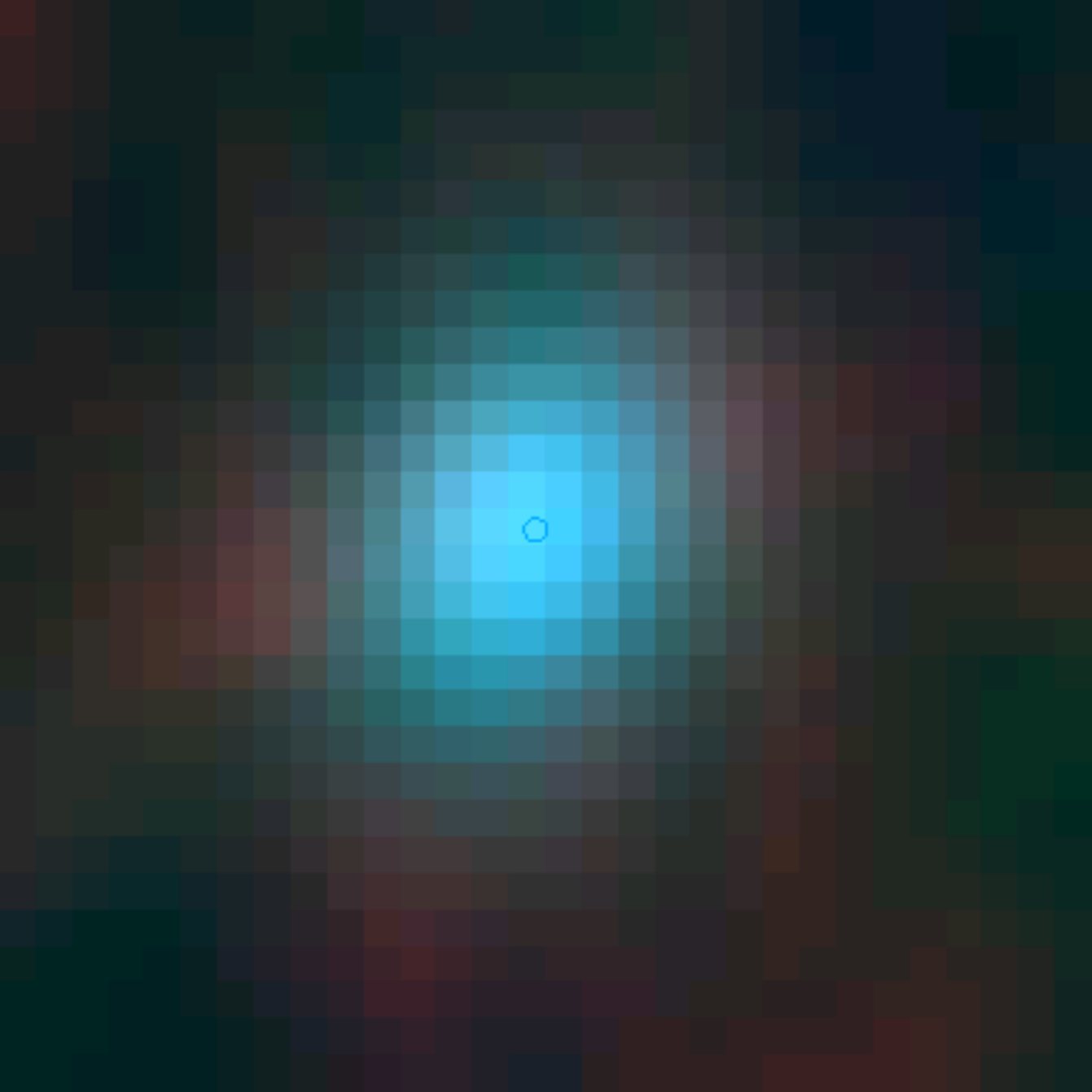}\hspace*{-.1em}
\includegraphics[width=2cm,height=2cm]{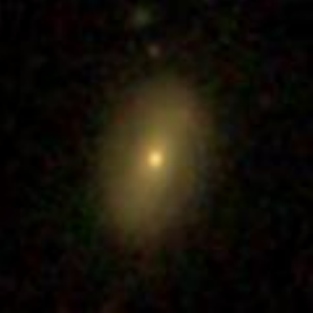}\hspace*{-.1em}\\[.6in]
\tiny NGC0070 & \tiny UGC00335NED01 & \tiny MCG+04-22-007\\
\includegraphics[width=2cm,height=2cm]{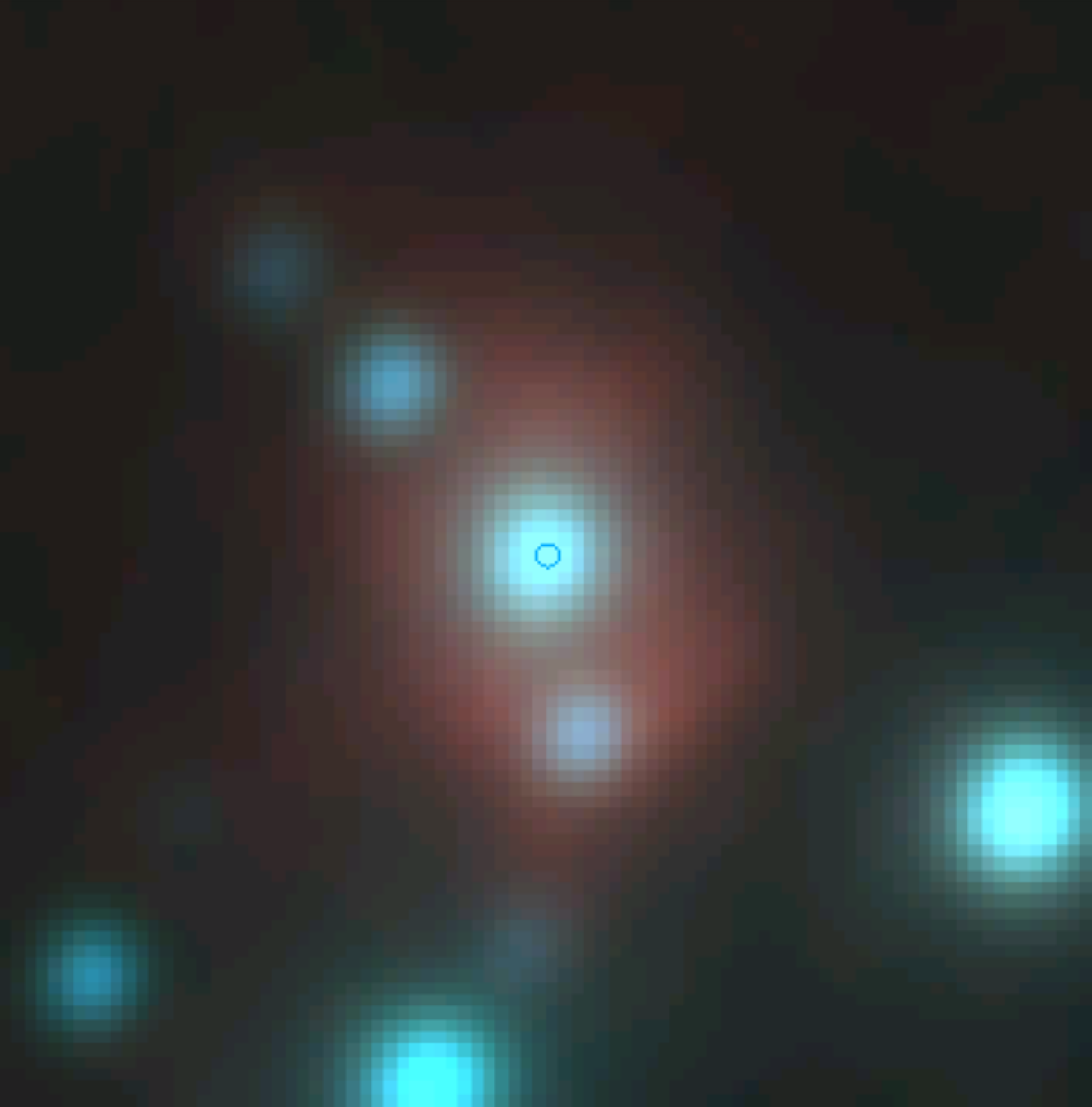}\hspace*{-.1em}
\includegraphics[width=2cm,height=2cm]{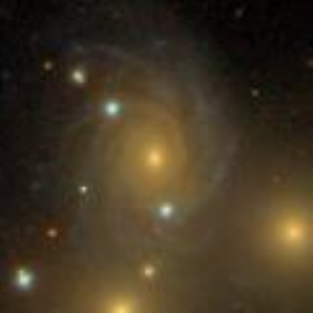}\hspace*{-.1em}&
\includegraphics[width=2cm,height=2cm]{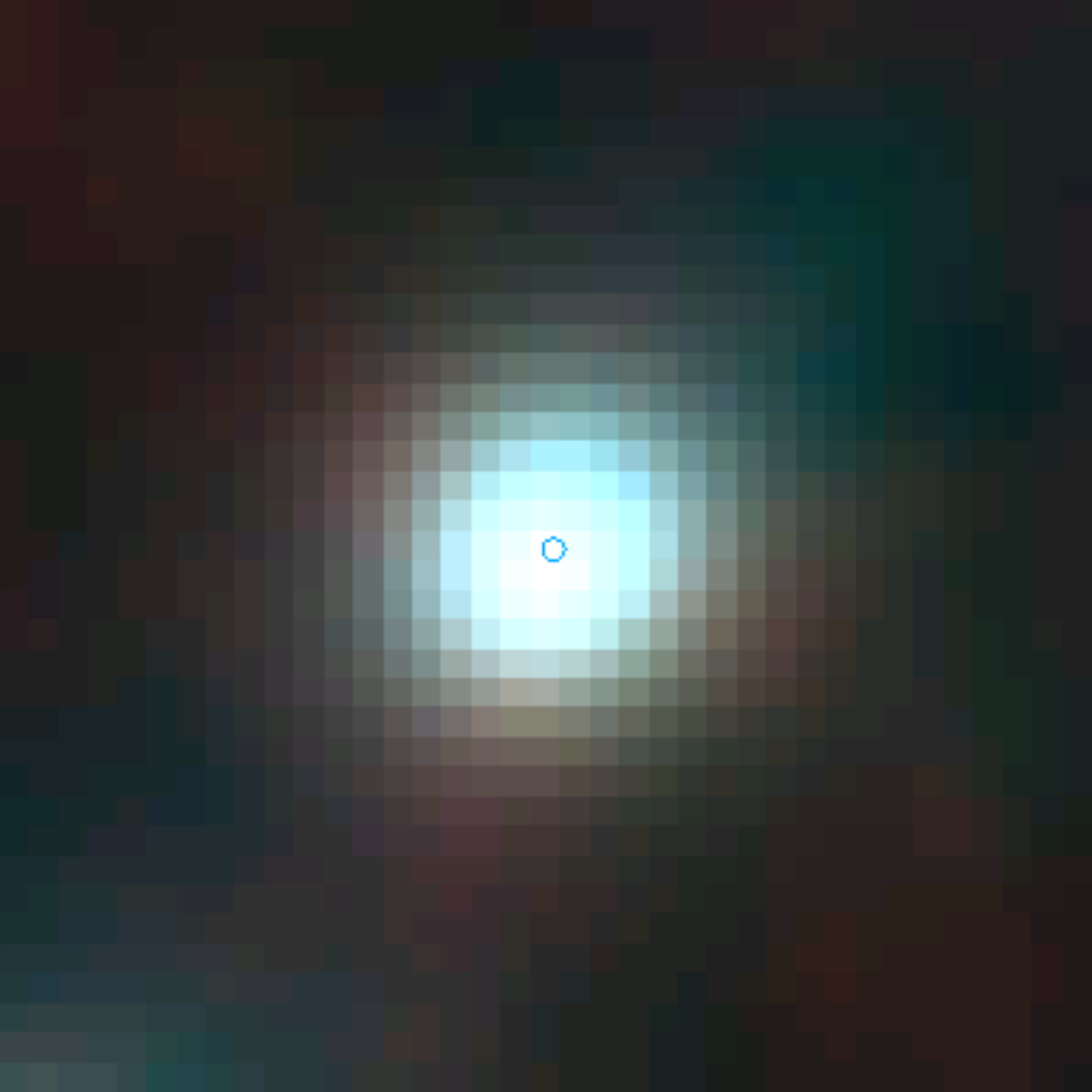}\hspace*{-.1em}
\includegraphics[width=2cm,height=2cm]{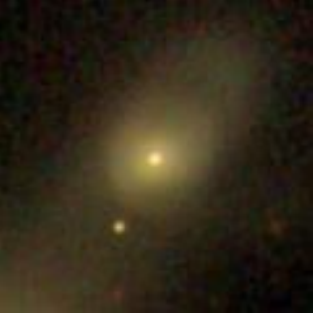}\hspace*{-.1em}&
\includegraphics[width=2cm,height=2cm]{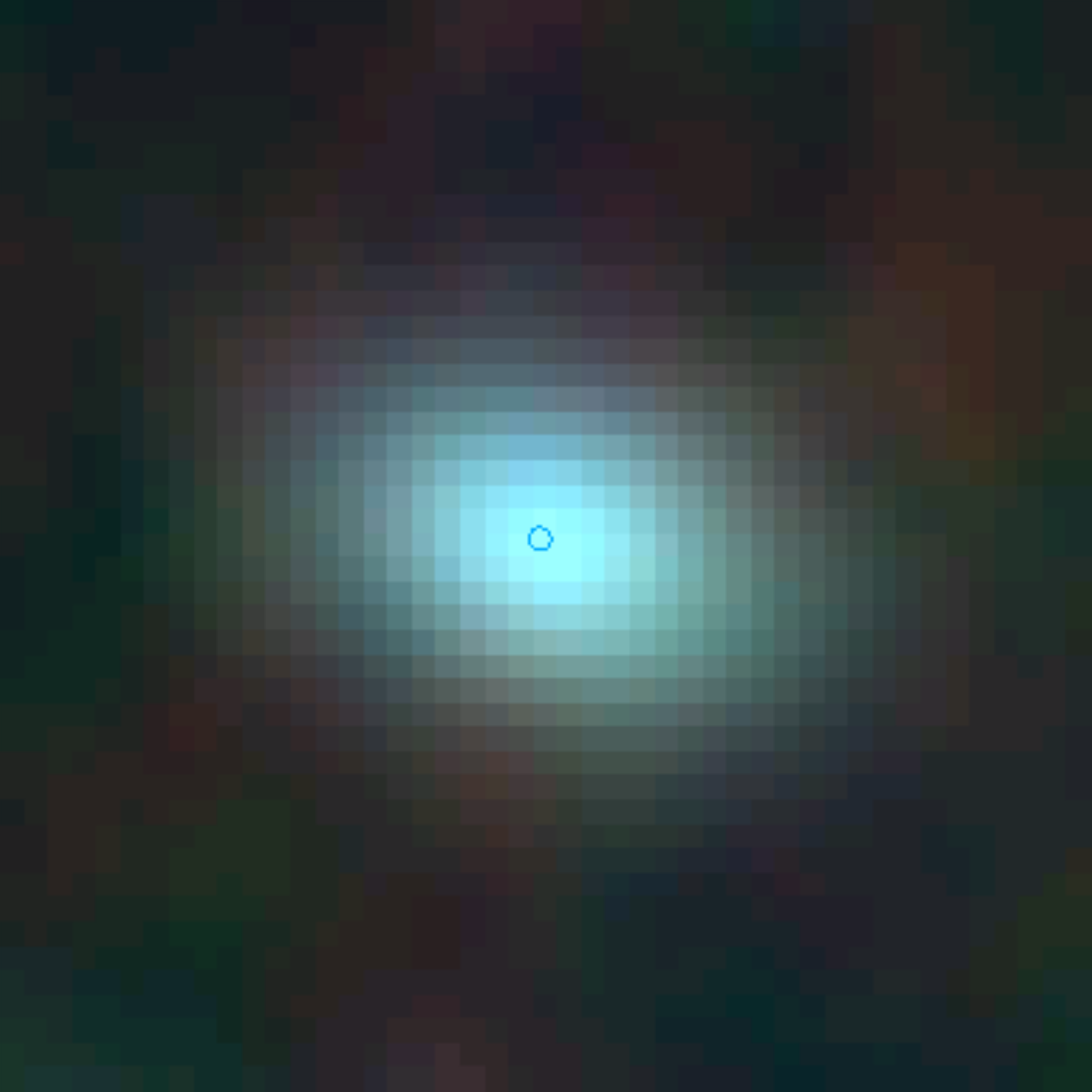}\hspace*{-.1em}
\includegraphics[width=2cm,height=2cm]{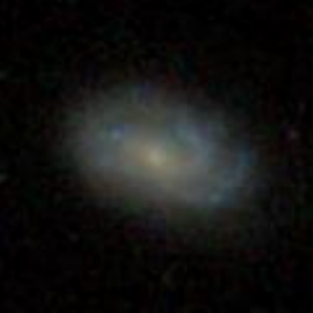}\hspace*{-.1em}\\[-.2in]
\tiny NGC2831 & \tiny NGC6296 & \tiny NGC4117\\
\includegraphics[width=2cm,height=2cm]{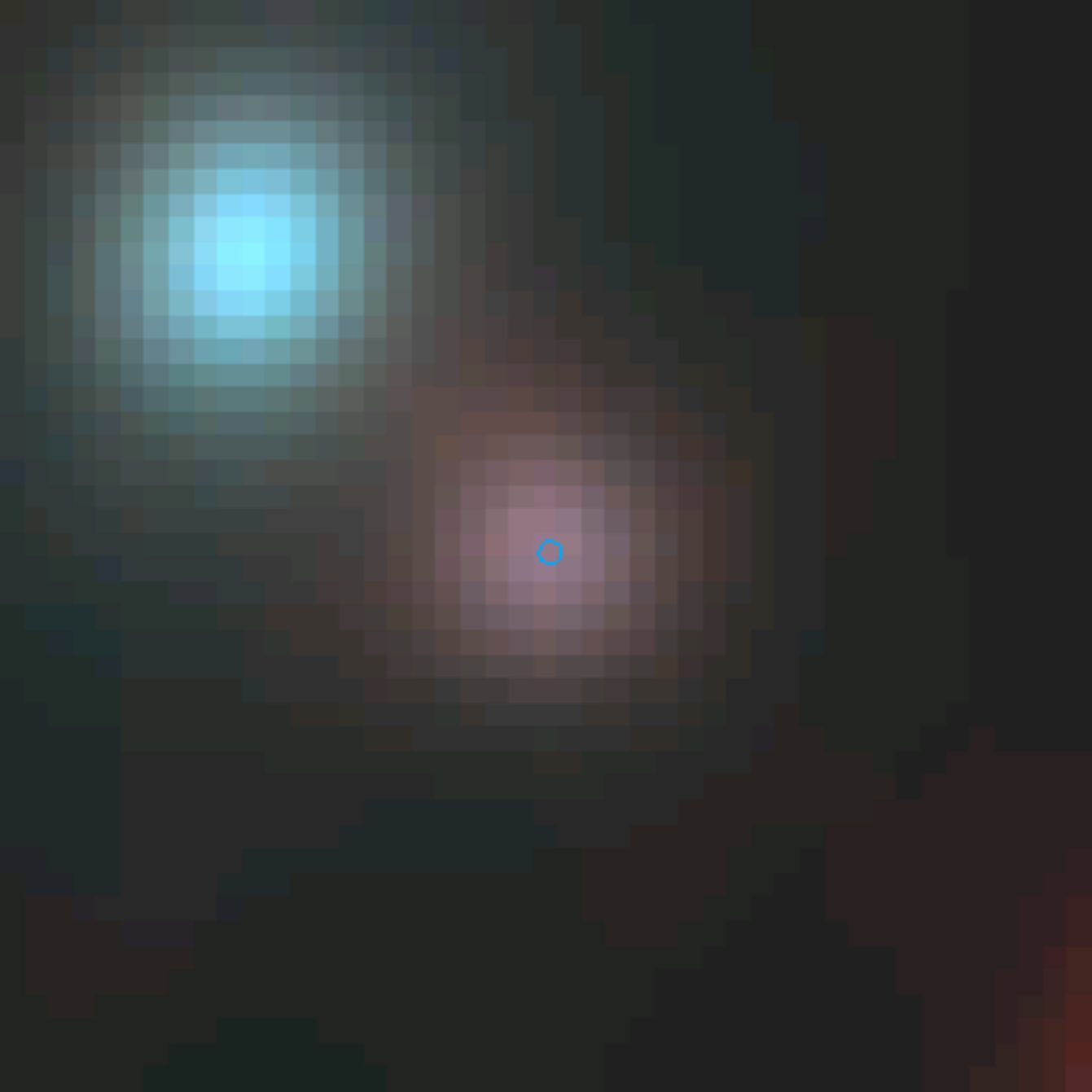}\hspace*{-.1em}
\includegraphics[width=2cm,height=2cm]{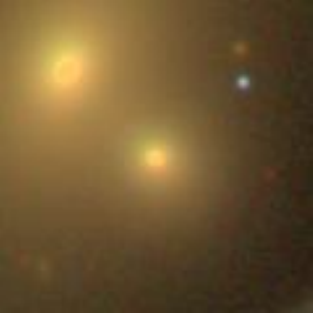}\hspace*{-.1em}&
\includegraphics[width=2cm,height=2cm]{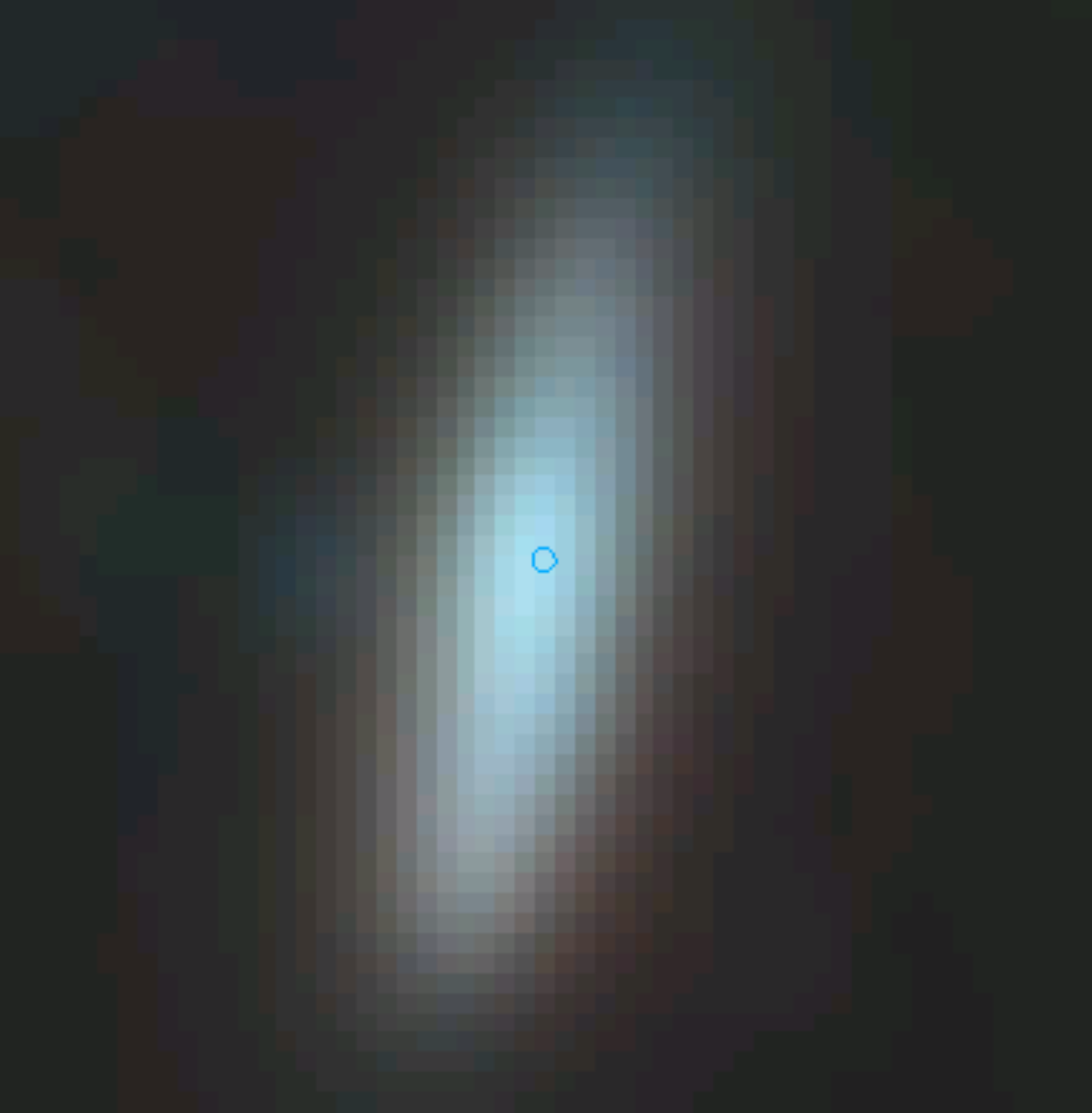}\hspace*{-.1em}
\includegraphics[width=2cm,height=2cm]{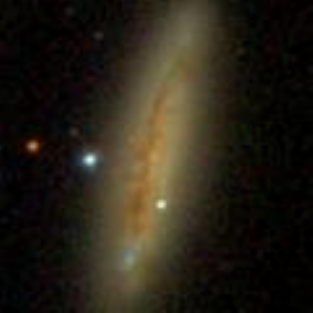}\hspace*{-.1em}&
\includegraphics[width=2cm,height=2cm]{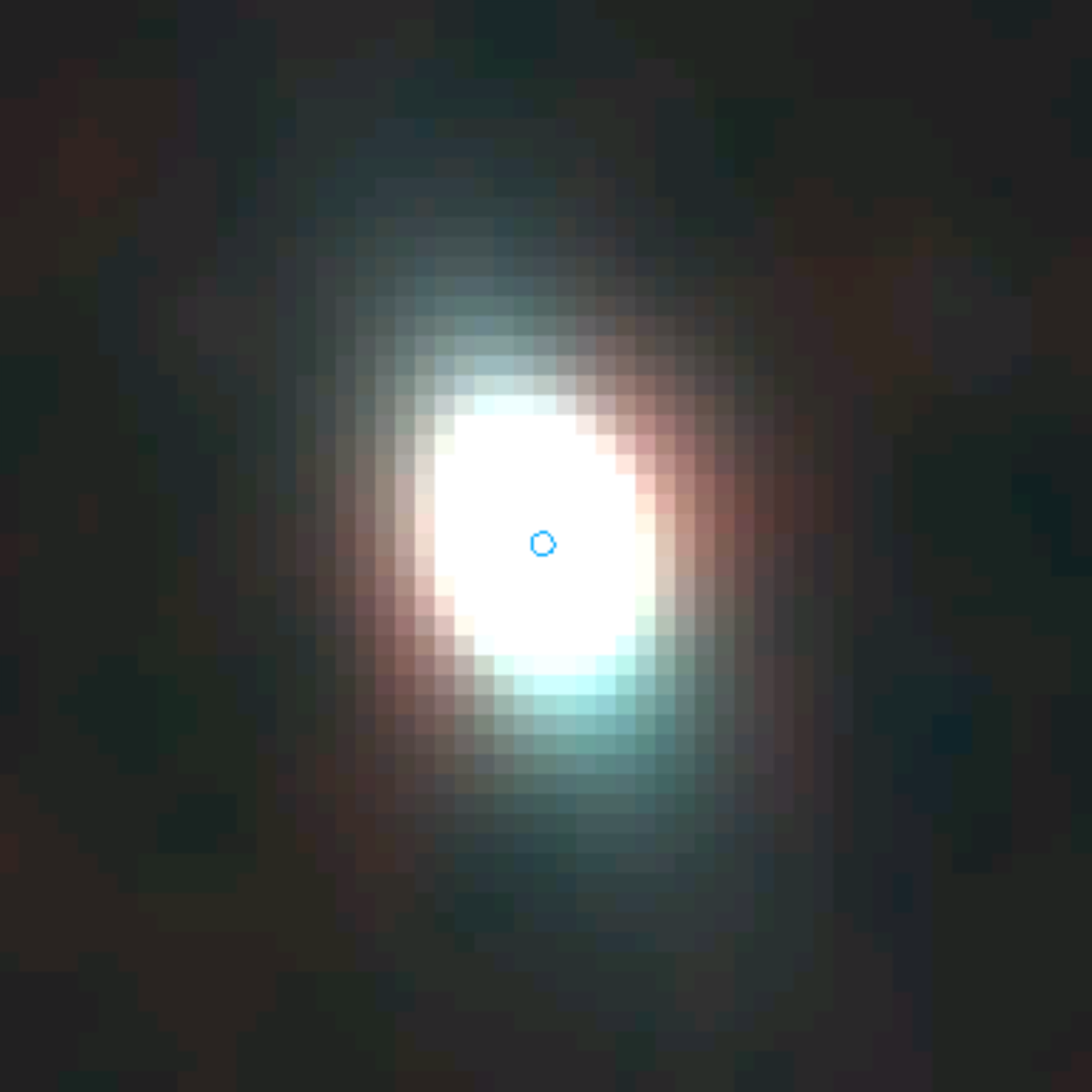}\hspace*{-.1em}
\includegraphics[width=2cm,height=2cm]{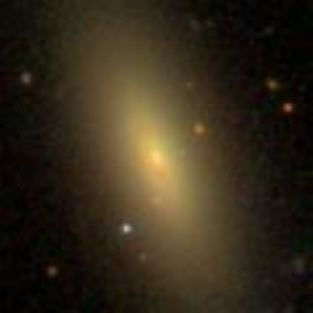}\hspace*{-.1em}\\[-.2in]
\tiny NGC4274 & \tiny NGC4314 & \tiny NGC4614 \\
\includegraphics[width=2cm,height=2cm]{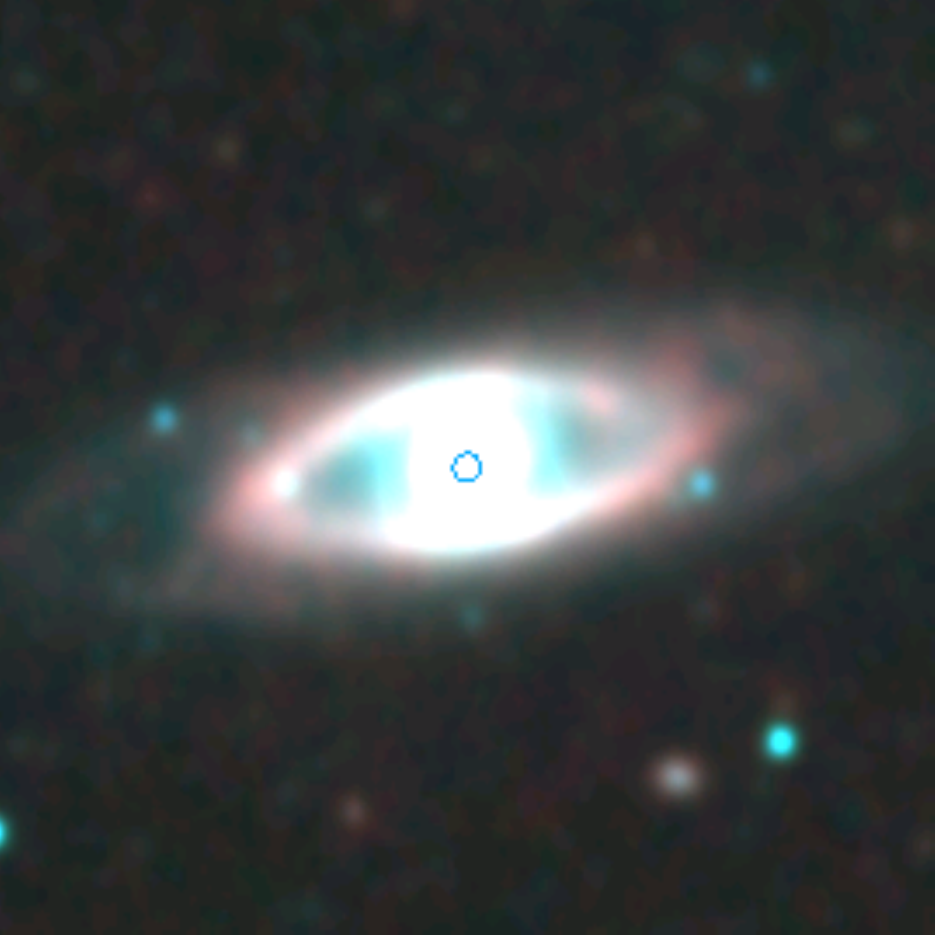}\hspace*{-.1em}
\includegraphics[width=2cm,height=2cm]{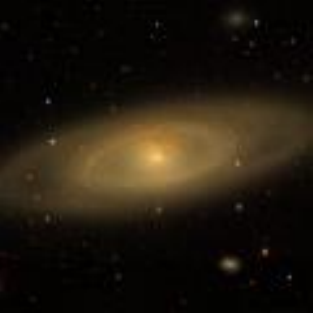}\hspace*{-.1em}&
\includegraphics[width=2cm,height=2cm]{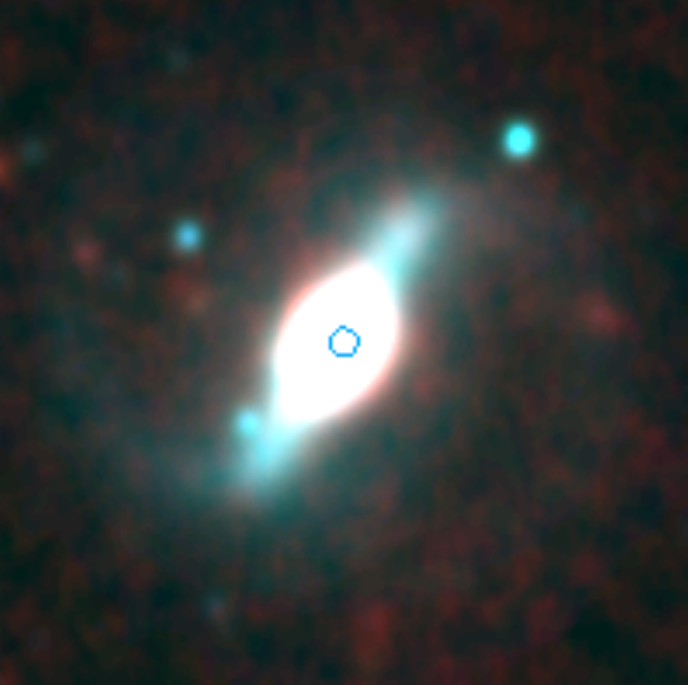}\hspace*{-.1em}
\includegraphics[width=2cm,height=2cm]{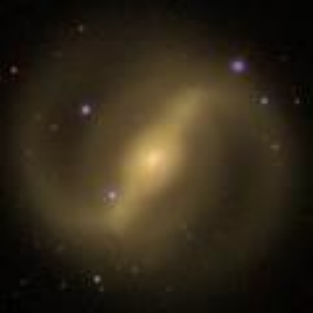}\hspace*{-.1em}&
\includegraphics[width=2cm,height=2cm]{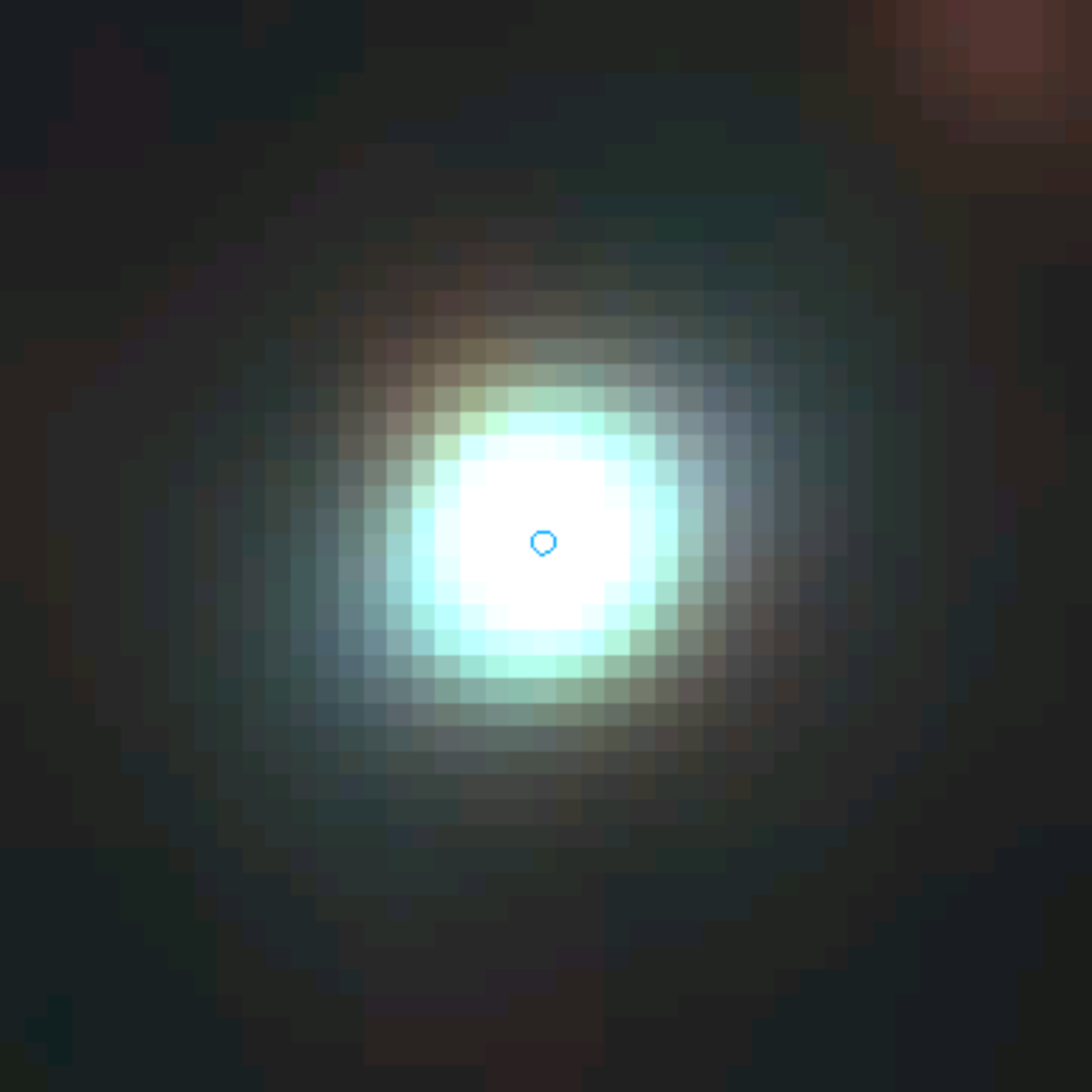}\hspace*{-.1em}
\includegraphics[width=2cm,height=2cm]{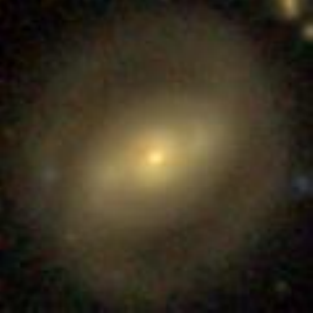}\hspace*{-.1em}\\[-.2in]
\tiny NGC6961\\
\includegraphics[width=2cm,height=2cm]{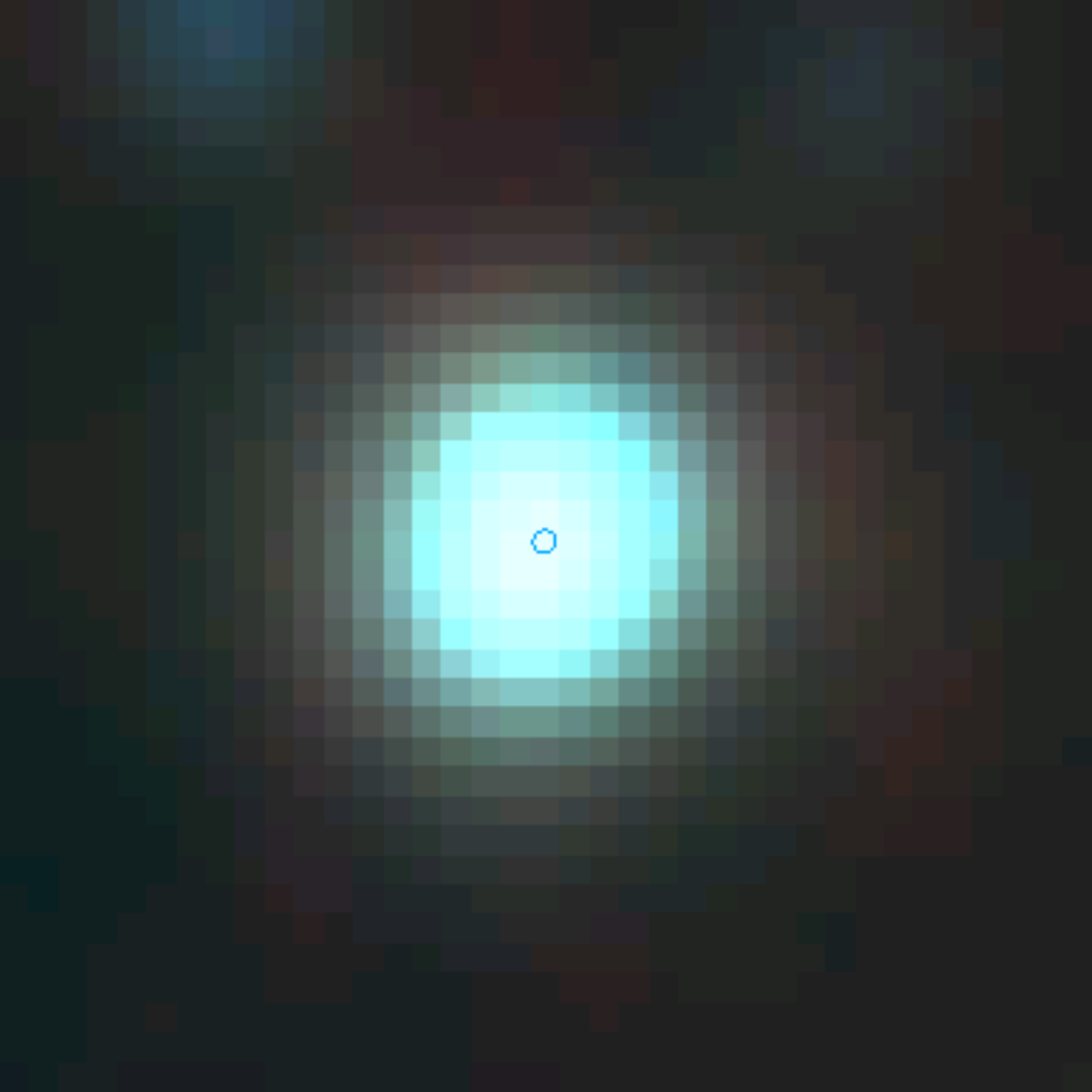}\hspace*{-.1em}
\includegraphics[width=2cm,height=2cm]{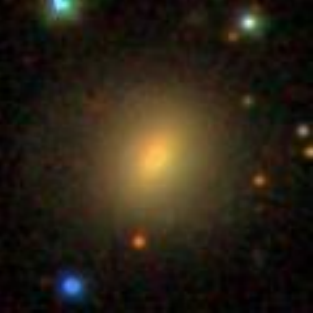}\hspace*{-.1em}\\
\end{longtable}
 \captionof{figure}{\singlespace Side-by-side comparison of \textit{Left:} Five-color [\textit{ugriz}] SDSS thumbnails and \textit{Right:} Three-color [$3.4\micron$ (blue), $4.6\micron$ (green), $12\micron$ (red)] composite WISE thumbnails of all the \textbf{$SNR > 2$} galaxies in the WISE canyon that fall within the SDSS footprint (31 galaxies). The size scale in each pair is identical and the scales range from $30''-360''$ over all thumbnails. The median PSF (\textit{r} band) of the SDSS cutouts is $1.3"$, while the typical PSF ($3.4\micron$, $4.6\micron$, $12\micron$ bands) for the WISE cutouts is $8.5"$. In the WISE images, red indicates active star formation while blue indicates quiescence, while the opposite is true in the SDSS images. }
  \addtocounter{table}{-1}
  \label{fig:thumbnails}

\end{center}

\singlespace

\raggedright
\parindent=1.8em

\subsection{WISE-SDSS Colorspace}
\begin{figure}[H]
\centering
\includegraphics[width=.6\textwidth]{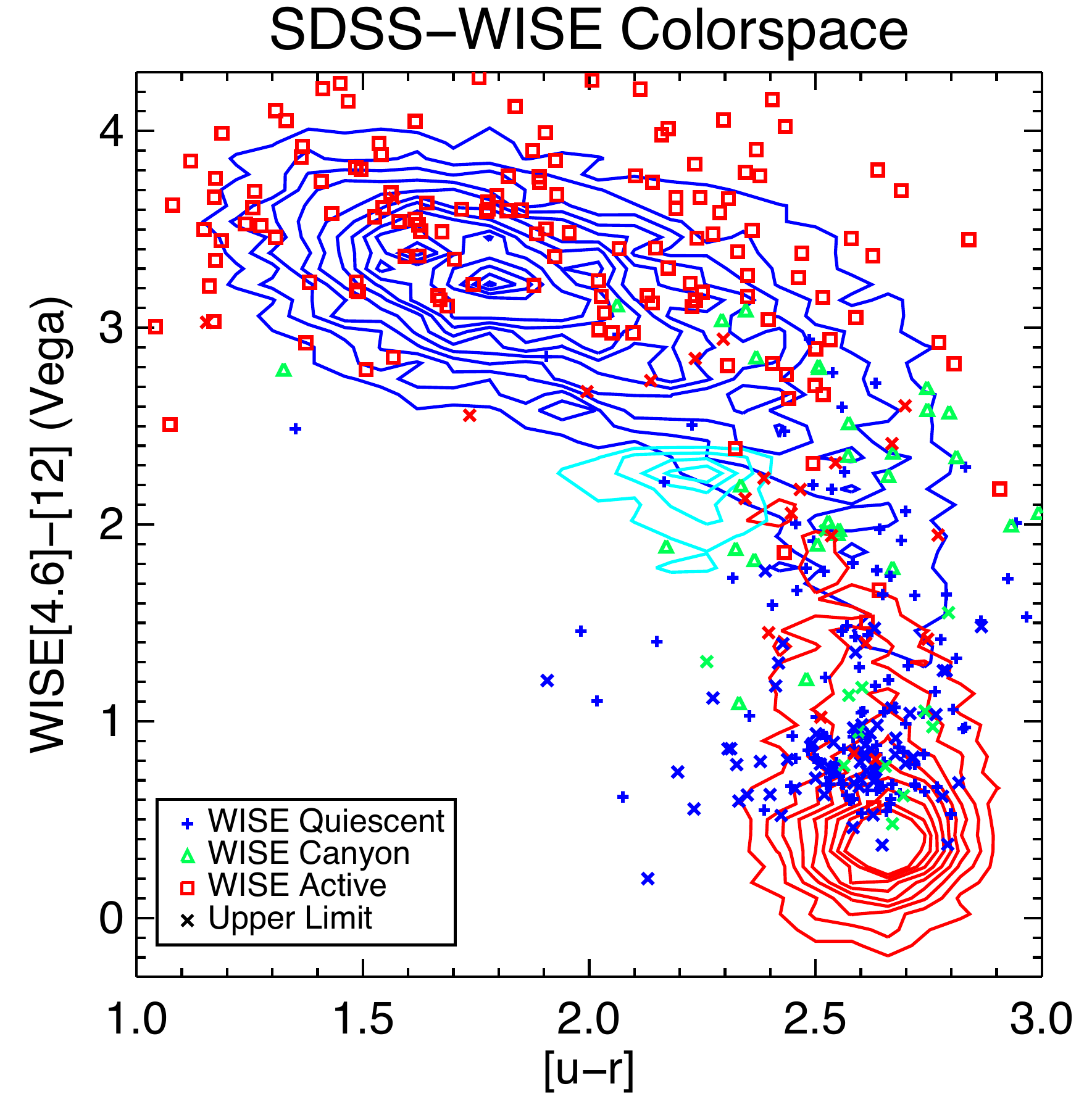}
\caption{Distribution of compact group galaxies in $\log[{\frac{\rm f_{12}}{\rm f_{4.6}}}]$ vs. [u-r] color, with contours (now uncorrected for intrinsic extinction) from \citet{alatalo14} overlaid. The galaxies are color-coded by their location in WISE color-color space; galaxies with an upper limit in either W3 or W4 (see Figure \ref{fig:colorspace_fullsamp}) are color-coded by the same classification but plotted with X's. We confirm a bimodality between morphologically classified early-type (red contours) and late-type (blue contours) galaxies, first seen for field galaxies in \citet{alatalo14}, but here for compact group galaxies, with a majority of our WISE canyon galaxies falling at the elbow of the two distributions. We also note a dearth of WISE canyon galaxies within the optical green valley (cyan contours).} 
\label{fig:wise_sdss}
\end{figure}

As seen in Figure \ref{fig:wise_sdss}, we also confirm a relation between $\log[{\frac{\rm f_{12}}{\rm f_{4.6}}}]$ vs. [u-r] color for compact group galaxies, a relation first noted for Galaxy Zoo sources in \citet{alatalo14}. They find a significant bifurcation between late-type (blue contours) and early-type (red contours) galaxies; the green valley galaxies (cyan contours) appear at the elbow of the two distributions. The \citet{alatalo14} contours are derived by plotting $\approx50,000$ field galaxies in this region of colorspace, which are characterized as either ``early", ``late", or ``green valley" objects in Galaxy Zoo. We plot the 407 galaxies from the full sample that fall within the SDSS footprint, and overlay the same contours as Figure 1f in \citet{alatalo14}, now uncorrected for intrinsic extinction. \citet{alatalo14} originally correct for intrinsic extinction using the $E(B-V)_{stars}$ from the \citet{oh11} OSSY catalog, which is then converted to an extinction measure at the rest wavelength of the galaxy. The OSSY database is derived from the SDSS DR7 spectral catalog, and only $\approx 2/3$ of our sample have SDSS spectral data; thus, we are not able to correct for intrinsic extinction following the methodology from \citet{alatalo14}. As a subset of our sample tends to be both dusty and actively star forming, correcting for intrinsic extinction would preferentially shift our active population towards lower [u-r] color, and the kink in the distribution would become more pronounced.

When comparing the photometry, we find an offset between the Galaxy Zoo [W2-W3] colors and our compact group [W2-W3] colors. The Galaxy Zoo magnitudes are composed largely of the w2gmag and w3gmag parameters in the ALLWISE source catalog, whose elliptical apertures are scaled from the 2MASS extended source catalog apertures. Specifically, the ALLWISE gmag aperture semi-major axes are 1.1 times the 2MASS XSC $K_s$ circular semi-major axes. When directly comparing the subsample of our compact group galaxies that are also Galaxy Zoo objects, we find a net offset of $\approx0.3-0.4$ magnitudes between the two samples. We attribute this offset to a difference in aperture type: the gmag elliptical apertures are determined using solely the $2.159\micron$ 2MASS $K_s$ image, while are custom $1-3\sigma$ contour apertures are determined using an averaged, $\lambda^{-1}$ weighted reference image composed of the $3.4\micron$, $4.6\micron$, and $12\micron$ images. Our offset is consistent with the findings of \citet{cluver14}, who quantify the difference between WISE gmag and isophotal photometry and determine a shift of $\approx 0.2-0.4$ magnitudes. The gmag apertures can also be contaminated by neighboring sources, which likely produces higher gmag values for faint sources when compared directly to the isophotal photometry \citep{cluver14}. Nevertheless, the shift is not significant with regards to our conclusions, and we note the same bifurcation between ``active" compact group galaxies (dominated by late-type Sc-Sd galaxies, see Figure \ref{fig:morphology}) and ``quiescent" compact group galaxies (dominated by early-type E/S0 galaxies). We also find that the WISE IR canyon galaxies cluster near the ``elbow" between the early-type (lower right red contours) and late-type (upper left blue contours) galaxy distributions, though they do tend to span a larger range in [W2-W3] color. We also note a dearth of canyon galaxies within the optical green valley (cyan contours), reconfirming that WISE infrared canyon galaxies \textit{do not fall in the optical green valley}. Instead, as argued in \citet{walker13}, we confirm that the majority of our WISE canyon galaxies tend to be optically redder than typical green valley objects in Galaxy Zoo, falling on or near the red sequence.

\section{Summary and Conclusions}
We perform custom photometry on a comprehensive sample of 163 compact groups and produce a catalog of 567 compact group galaxies in all four WISE bands. Within this sample, we present the identification of 37 moderately star-forming WISE canyon galaxies which reliable photometry that lie in a distinct region of WISE $\log[{\frac{\rm f_{12}}{\rm f_{4.6}}}]$ vs. $\log[{\frac{\rm f_{22}}{\rm f_{3.4}}}]$ color-color space. Canyon galaxies are severely under-represented in the previous \spit sample, and the small number statistics prevent an understanding of the dominant evolutionary mechanisms at play. With this enlarged sample, it should be possible to begin statistically characterizing the properties of the WISE canyon galaxies, and to examine what processes may be causing their accelerated evolution. We also note that, despite the WISE canyon region being less pronounced than the \textit{Spitzer} IRAC canyon, we can also utilize this WISE colorspace to separate galaxies dominated by stellar light from galaxies dominated by PAH and thermal emission. The creation of a photometric catalog could aid future studies seeking to characterize the properties of these unique galaxies and their accelerated evolution. 

We also note distinct trends between WISE classification [quiescent, canyon, active] and both stellar mass and star formation rate. Though the classes span approximately the same range of stellar masses, the WISE quiescent class tends to have systematically higher stellar masses than the canyon class, while the canyon class has systematically higher stellar masses than the active class. The opposite is predictably true for star formation rate: though the classes overlap in SFR space, the active class has the highest median SFR, followed by the canyon class and the quiescent class.

Similar to \citet{alatalo14}, we find that compact group galaxies also exhibit a bimodality between early and late type galaxies in WISE $\log[{\frac{\rm f_{12}}{\rm f_{4.6}}}]$ vs. $\log[{\frac{\rm f_{22}}{\rm f_{3.4}}}]$ color-color space, with a plurality of WISE narrow canyon galaxies being of type Sa-Sbc. However, we note that the canyon between spiral and bulge-dominated galaxies does not correspond to the optical green valley and confirm the results from \citet{walker13} that a majority of WISE canyon galaxies tend to be optically redder. 

Ultimately, this statistical sample provides a means to address how galaxies are transformed in the high-density and dynamically-evolving compact group environment, and consequently what impact this key environment has on galaxy evolution and hierarchical formation. A number of processes could be responsible for this accelerated evolution---possibilities include briefly enhanced star formation due to a tidal inflow of gas; rapidly quenched star formation to the tidal removal of gas; or strangulation if cool filaments of gas are not able to refuel the system.  Determining the relative importance of these and other mechanisms is essential to understanding how the gas is processed in these groups, which has a direct impact on the resulting merger products. 



\section{Acknowledgments}
This publication makes use of data products from the Wide-field Infrared Survey Explorer, which is a joint project of the University of California, Los Angeles, and the Jet Propulsion Laboratory/California Institute of Technology, funded by the National Aeronautics and Space Administration

Funding for the Sloan Digital Sky Survey IV has been provided by
the Alfred P. Sloan Foundation, the U.S. Department of Energy Office of
Science, and the Participating Institutions. SDSS-IV acknowledges
support and resources from the Center for High-Performance Computing at
the University of Utah. The SDSS web site is www.sdss.org.

Catherine S. Zucker would like to sincerely thank the anonymous referee for their thorough feedback throughout the review process. She would also like to thank the Virginia Space Grant Consortium and the University of Virginia College of Arts and Sciences for support. 

Sarah C. Gallagher thanks the Natural Sciences and Engineering Research Council of Canada for support. 




\bibliographystyle{apj}
\bibliography{Zucker_CG_Wise_Revised}

\section{Appendix}
\subsection{Canyon Bounds}
Quantitatively, we define our canyon zone to be where: \[\rm ywise \ge (0.01866-1.885\times xwise) \land ywise \le (0.7844-1.885\times xwise)\] 
where xwise=$\log[{\frac{\rm f_{22}}{\rm f_{3.4}}}]$ color and ywise=$\log[{\frac{\rm f_{12}}{\rm f_{4.6}}}]$ color

\setlength{\tabcolsep}{1pt}
\tabletypesize{\scriptsize\tiny}
\begin{deluxetable}{ccccccccccc}
\tablecaption{{WISE Galaxies from Full Sample} \label{tab:canyongals}}
\tablecolumns{11}
\tablewidth{0pc}
\tablehead{
\colhead{Group} & \colhead{Galaxy} & \colhead{W1} & \colhead{W2} &
\colhead{W3} & \colhead{W4} & \colhead{Morphology} & \colhead{De Vaucouleurs} &\colhead{WISE} & \colhead{Photometry} &\colhead{Upper}\\
\colhead{} & \colhead{} & \colhead{mJy} & \colhead{mJy} & \colhead{mJy} & \colhead{mJy} & \colhead{} & \colhead{number*} & \colhead{Class**} & \colhead{type***} & \colhead{Limit****}}
\startdata
HCG01&a&3.43$\pm$0.08&2.04$\pm$0.05&7.59$\pm$0.22&22.50$\pm$1.45&Sc&6.0&Active&3&No\\
HCG01&b&4.06$\pm$0.09&2.15$\pm$0.05&1.96$\pm$0.14&6.92$\pm$1.02&E&-3.5&Canyon&3&No\\
HCG01&c&3.54$\pm$0.06&1.82$\pm$0.04&0.67$\pm$0.02&3.42$\pm$6.54&E-S0&-3.4&Quiescent&2&Yes\\
HCG01&d&0.95$\pm$0.05&0.49$\pm$0.03&0.55$\pm$0.02&4.86$\pm$4.78&S0&-1.9&Active&2&Yes\\
HCG02&a&9.42$\pm$0.09&6.11$\pm$0.07&33.18$\pm$0.17&87.30$\pm$1.52&SBd&8.0&Active&1&No\\
HCG02&b&11.90$\pm$0.05&8.49$\pm$0.03&79.40$\pm$0.07&373.56$\pm$7.72&S0-a&-1.1&Active&2&No\\
HCG02&c&5.56$\pm$0.06&2.94$\pm$0.01&13.16$\pm$0.19&8.14$\pm$1.66&SABc&4.9&Active&2&No\\
HCG03&a&6.92$\pm$0.04&4.05$\pm$0.02&17.24$\pm$0.02&22.23$\pm$1.44&Sc&4.6&Active&2&No\\
\enddata
\vspace{-0.4cm} 
\tablecomments{*The morphological types and De Vaucouleurs numbers are from the HyperLeda database \citep{paturel03a,paturel03b}, which accumulates morphologies from available publications and compares and combines them to determine an optimal De Vaucouleurs number and classification for each galaxy. **See \S3.2 and Figure \ref{fig:colorspace_fullsamp} for explanation of WISE classification scheme [Quiescent, Canyon, Active]. ***See \S2.2 for description of photometry types. ****This column indicates whether there is an upper limit in a single band (either W3 or W4); galaxies with upper limits in more than one band are excluded from the full sample. In cases where the galaxy has an upper limit, the error in that band is set to null in the ALLWISE source catalog, and we quote the ratio of $\frac{S}{SNR}$ as the uncertainty. }
\end{deluxetable}

\end{document}